\documentclass[aps,pre,twocolumn,groupedaddress,floatfix]{revtex4-2}
\usepackage{graphicx}
\newcommand{\dd}{\mathrm{d}}
\usepackage{amsmath}
\DeclareMathOperator{\atantwo}{atan2}
\usepackage{amsfonts}
\usepackage{RefJ}
\usepackage{subcaption}

\usepackage[normalem]{ulem}
\usepackage[x11names]{xcolor}

\begin{document}

\title{BxC: a swift generator for 3D magnetohydrodynamic turbulence}

\author{Jean-Baptiste Durrive$^1$}
\author{Madhurjya Changmai$^1$}
\author{Rony Keppens$^1$}
\email[]{rony.keppens@kuleuven.be}
\author{Pierre Lesaffre$^2$}
\author{Daniela Maci$^1$}
\author{Georgios Momferatos$^3$}

\affiliation{$^1$ Centre for mathematical Plasma Astrophysics, Department of Mathematics, KU Leuven, 3001 Leuven, Belgium}
\affiliation{$^2$ Laboratoire de Physique de l'Ecole normale sup\'erieure, ENS, Universit\'e PSL, CNRS, Sorbonne Universit\'e, Universit\'e de Paris, 75005 Paris, France}
\affiliation{$^3$ Environmental Research Laboratory, National Center for Scientific Research Demokritos, Agia Paraskevi, 15341, Greece}

\date{\today}

\begin{abstract}
Magnetohydrodynamic turbulence is central to laboratory and astrophysical plasmas, and is invoked for interpreting many observed scalings. Verifying predicted scaling law behaviour requires extreme-resolution direct numerical simulations (DNS), with needed computing resources excluding systematic parameter surveys. We here present an analytic generator of realistically looking turbulent magnetic fields, that computes 3D ${\cal{O}}(1000^3)$ solenoidal vector fields in minutes to hours on desktops. Our model is inspired by recent developments in 3D incompressible fluid turbulence theory, where a Gaussian white noise vector subjected to a non-linear transformation results in an intermittent, multifractal random field. Our $B\times C$ model has only few parameters that have clear geometric interpretations. We directly compare a (costly) DNS with a swiftly $B\times C$-generated realization, in terms of its (i) characteristic sheet-like structures of current density, (ii) volume-filling aspects across current intensity, (iii) power-spectral behaviour, (iv) probability distribution functions of increments for magnetic field and current density, structure functions, spectra of exponents, and (v) partial variance of increments. The model even allows to mimic time-evolving magnetic and current density distributions and can be used for synthetic observations on 3D turbulent data cubes.
\end{abstract}

\maketitle

\section{Introduction}

Fluids and magnetic fields are usually turbulent, and researchers often need to model and analyze turbulent data. Since fully nonlinear, turbulent, analytic solutions to the Navier-Stokes (hydro) or the magnetohydrodynamic (MHD) equations are unavailable, the most common tool to construct realistic models is by means of direct numerical simulations (DNS), which are -- unfortunately -- extremely expensive resource-wise~\footnote{DNSs serve to simulate all processes up to the numerical resolution. This may also use Reynolds Averaged Navier-Stokes prescriptions for following the time-averaged fields, but is distinct from Large Eddy Simulations where large eddies are fully solved for while eddies below a cut-off are modelled.}. This led to the creation of online turbulence databases, \citep[e.g.][for the Johns Hopkins Turbulence Database]{Li2008} where selected snapshots of isotropic hydro turbulent fields up to $8192^3$ size, or $1024^3$ incompressible MHD states, are stored for web-based access. To date, DNS models of increasingly larger size provide the only means to verify theoretical scaling laws, which for MHD in particular, are still subject of lively contemporary debate \citep[e.g. see][]{Alex2020}. MHD turbulence, especially in 3D incompressible settings, is discussed in many modern textbooks \citep[e.g.][]{GKP2019,Biskamp2009,Galtier2016}, and these invariably emphasize its scaling and shape in power spectra, and its typical current-sheet dominated visual appearance.

In the quest for finding `exact' solutions to the incompressible Navier-Stokes equations, \cite{ChevillardEtAl10} suggested an explicit, concise, and yet efficient, analytical expression for a random field which shares many properties of experimental and numerical incompressible hydrodynamical turbulence \citep[see also][]{ChevillardEtAl11, ChevillardEtAl12, ChevillardEtAl13, ChevillardHDR, PereiraEtAl16, PereiraEtAl18, ReneuveChevillard20, ApolinarioEtAl22}. Intermittency (i.e. non-Gaussianity) in this model stems from the fact that the random field is constructed as products, i.e. a non-linear transformation of Gaussian white noises. For this reason this approach belongs to the mathematical field called `Gaussian multiplicative chaos', first formalized by \cite{Kahane85}.
In order to build similarly parametric models for astrophysical environments, \cite{DurriveEtAl20} recently suggested an extension of the aforementioned model to magnetized fluids, mimicking MHD turbulence. In these constructed random fields, their statistics are controlled by a couple of free, physically motivated, parameters.

The approaches above have a threefold ambition: The random fields must (i) resemble real data as much as possible, (ii) be physically motivated, and (iii) be as numerically efficient as possible, to be worthwhile compared to DNSs.
They are useful in many ways, e.g. to quickly generate synthetic data (effective, `surrogate', models), to characterize turbulent data with few parameters for observers or experimentalists, and for constructing non-trivial (i.e. with at least self-similar and small-scale structuring) initial conditions for DNSs.
In the currently latest HD \cite{ChevillardEtAl10} or MHD \cite{DurriveEtAl20} efforts of this kind, objectives (ii) and (iii) are satisfyingly fulfilled, as the models are constructed from the physics of vortex stretching and flux tube shearing, and numerically they are several hundreds of times less resource consuming than DNS. As for objective (i) to resemble real turbulent data, in the hydrodynamical case all efforts have focused on the statistics of the fields, but not on the shape of the structures. Hence, while many statistical properties of the random incompressible velocity fields are fairly realistic, their 3D visualizations are far less convincing.

We here present a path to solve this problem, i.e. to build very efficiently (objective (iii)) random fields that visually resemble DNS results (objective (i)). We do this here directly for the MHD case, where the challenge is to get both current and magnetic field vector quantities behave in DNS-like fashion. We name our model $B\times C$, standing for `magnetic fields from multiplicative chaos'.
Our reasoning is purely geometric, in the sense that we motivate our parametrized transformation mostly from getting visual correspondence with 3D turbulent magnetic vector fields. In practice, these parameters also relate, in a yet-to-be-quantified fashion, to the physical processes of vortex stretching and shearing (as we will `deform' spiral patterns based on gradient fields). The geometric parameters also are inspired by, and impact on, the statistical properties of the 3D turbulent states, and we provide various quantitative comparisons further on, notably in terms of energy spectra.

Incidently, it is straightforward to also adapt our model to the 2D case, by starting with the well-known 2D Biot-Savart's law and keeping the eddy modeling two-dimensional as in section \ref{sec:isolated_eddy}. The interested reader may have a look at for example \citep[][]{ReneuveChevillard20} who do work with fractional Gaussian fields (see definition below) in 2D. However, the strength of our model lies on its 3D nature, since 2D DNSs are fairly cheap to run and $B \times C$ is an interesting complimentary tool to DNSs only in the 3D case.

The paper is organized as follows. In the first part of the paper, we detail the construction of our model. After giving some background, we construct a formula mimicking an isolated eddy in 2D, as a set of constant-curvature spirals swirling around a single point. Then, in an efficient single mathematical step, we extend this formula to 3D sheets, with non-uniform curvature, randomly distributed throughout space. We also expose how to straightforwardly emulate a time evolution of our turbulent magnetic field. In the second part of the paper, we show an example of a 3D vector magnetic field and its current density built with our model, and compare them to a modern DNS result. The comparison is performed in multiple ways, inspecting several visual aspects and by means of quantitative statistical tools.

\section{Magnetic field construction}

\subsection{Preliminaries}

Biot-Savart's law expresses a magnetic field $\vec{B}$ in terms of its current density $\vec{j}$ as the convolution
\begin{equation}
\vec{B} = N_B \int_{\mathbb{R}^3} \frac{\vec{j} \times \vec{r}}{r^3} \dd V,
\label{def:BS}
\end{equation}
where $N_B \equiv \mu_0/4 \pi$, with $\mu_0$ the vacuum permeability. Inside all integrals we use the usual short-hand notations $\vec{r} \equiv \vec{x} - \vec{y}$ and $r \equiv |\vec{r}|$, not to be confused later with the 2D $(r,\theta)$ polar $r$-coordinate.

The basic structure of the models in \cite{ChevillardEtAl10} and \citep{DurriveEtAl20} is the modified version of Biot-Savart's law
\begin{equation}
\vec{B} = N_B \int_{r \leq L} \frac{\vec{c} \times \vec{r}}{(r^2+\eta^2)^{h}} \dd V.
\label{def:B}
\end{equation}
Compared to (\ref{def:BS}), the integration region is restricted to a ball of radius $L$, the kernel's fixed $r^{-3}$ power-law behavior is set to vary freely with a power $h$, and the kernel's singularity at $r=0$ is regularized with $\eta$, so that parameters $L, h$, and $\eta$ respectively serve to control the large-scale cut-off, the slope and the small-scale cut-off of the power spectrum of $B \equiv |\vec{B}|$. The normalizing constant $N_B$ will be used to control the total energy of the field (moving vertically the power spectrum, cf section \ref{sec:power_spectra)}).
Finally, we write $\vec{c}$ instead of $\vec{j}$, because due to the above modifications, $\vec{c}$ in (\ref{def:B}) is not exactly the current density $\vec{j}$ anymore. The strategy is to first construct $\vec{c}$, then deduce $\vec{B}$ through (\ref{def:B}), and only then deduce $\vec{j} \equiv \vec{\nabla} \times \vec{B}$ from $\vec{B}$.
Most importantly, the form (\ref{def:B}) guarantees $\vec{B}$ to be divergence-free, for any $\vec{c}$ (so $\vec{\nabla} \cdot \vec{c} = 0$ is not required) as long as $\eta$ is large enough for the field to be smooth on small scales such that gradients are well approximated \citep[e.g.][]{PereiraEtAl16}. At the same time we better take $\eta$ to be small, to have a large inertial range, and we take as a trade-off between these two constraints $\eta = 3/N$ at a resolution~$N$.

The core of this model is to choose a relevant $\vec{c}$.
Hereafter we call $\widetilde{s}$ a Gaussian white noise vector, the tilde symbol reminding its random nature and `s' standing for `seed'. The three components of $\widetilde{s}$ are Gaussian white noises, independent of one another, zero-averaged, and with unit variance.
The simplest idea takes $\vec{c}$ equal to $\widetilde{s}$ and (\ref{def:B}) reduces to
\begin{equation}
\vec{R} \equiv N_R \int_{r \leq L_R} \frac{\widetilde{s} \times \vec{r}}{(r^2+\eta_R^2)^{h_R}} \dd V,
\label{def:R}
\end{equation}
a field referred to as a fractional Gaussian field (fGf) \citep{ReneuveChevillard20}. We renamed it to $\vec{R}$ for reasons that will become clear further on, and added subscripts $R$ to the parameters in (\ref{def:R}) as they will have different numerical values than in (\ref{def:B}).
Now, magnetic fields in nature are non-Gaussian (intermittent), while $\vec{R}$ is still a Gaussian field because the Biot-Savart operation (\ref{def:R}) is a linear transformation (namely a convolution) on a Gaussian field (namely $\widetilde{s}$).
To build an intermittent model, we must find a non-linear transformation on $\widetilde{s}$, which is the purpose of the two following sections.

\subsection{Designing an isolated eddy}
\label{sec:isolated_eddy}

\begin{figure*}
\includegraphics[scale=0.36]{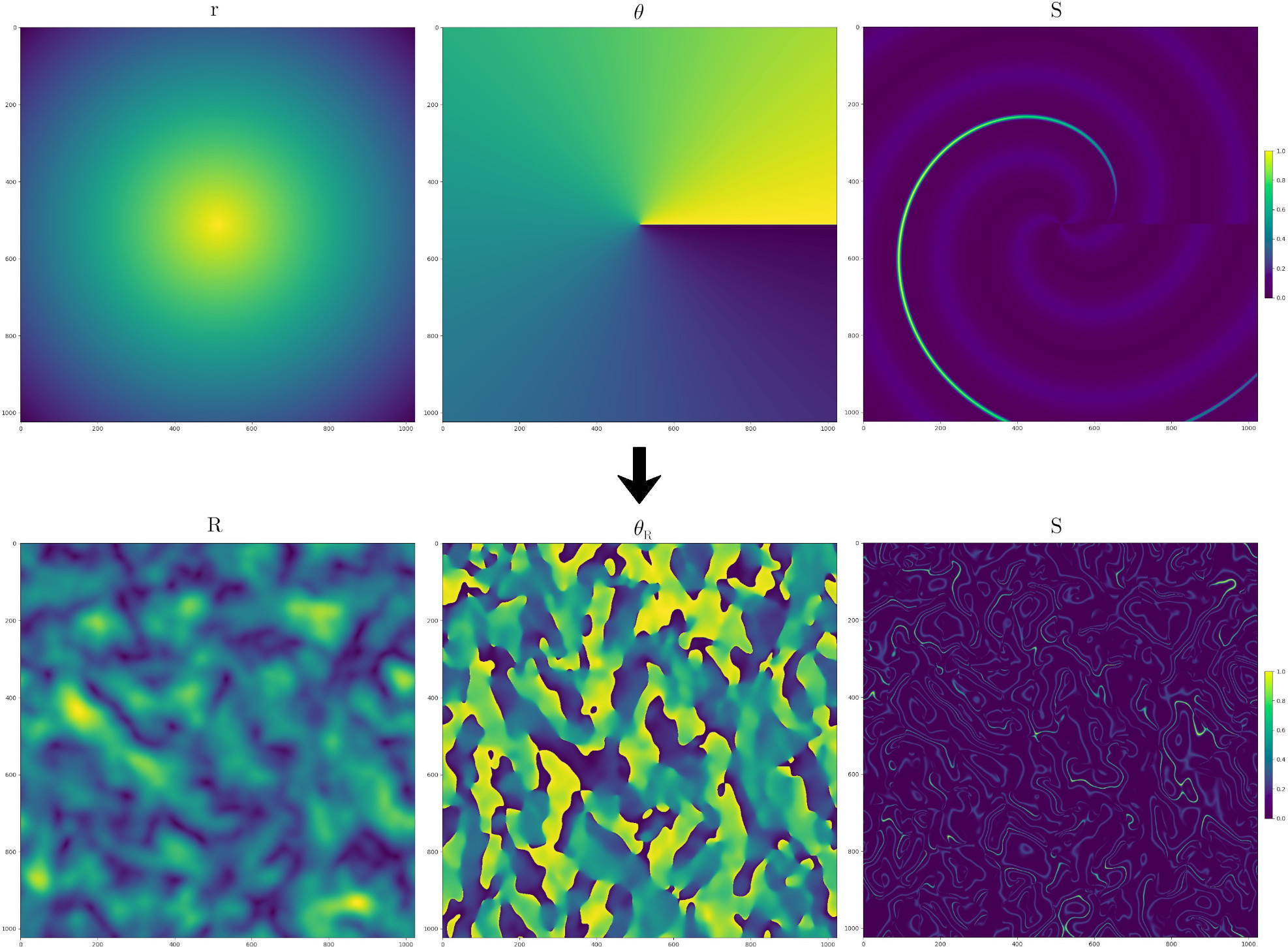} 
\caption{\label{fig:r_theta_S_2D_and_3D} (Color online) From deterministic spirals in 2D to random sheets in 3D. \textit{(Top row)} 2D setup: Using the deterministic $r_p$ field in the left panel and the $\theta$ field in the middle panel, we construct with (\ref{def:S}) the spiral-shaped field $S$ on the right. This $S$ could mimic an isolated eddy. \textit{(Bottom row)} 3D setup, generalizing the top row: Using the random $R$ field (norm of (\ref{def:R})) in the left panel and the $\theta$ field (\ref{def:theta_with_R}) in the middle panel, we construct similarly the field $S$ with swirling sheets on the right. This $S$ is used in (\ref{def:B}) to mimic a distribution of eddies.}
\end{figure*}

A characteristic feature of fully developed MHD turbulence is the omnipresence of current density sheets swirling around throughout space. This is -- in a sense -- the MHD counterpart of the vortex tubes from hydrodynamics, where eddies of varying sizes communicate nonlinearly.
In the simplest non-trivial model to mimic a swirling current sheet, we are led to the Archimedean spiral (such spiral shapes may represent the outcome of some underlying physical process, such as a Kelvin-Helmholtz roll-up pattern in a shear flow \citep[e.g.][]{GKP2019}; This Archimedean spiral also returns as the Parker spiral of interplanetary magnetic field \citep{Parker58} in an important historical model for solar wind magnetic fields). The latter is described in the 2D plane by the polar equation
\begin{equation}
r(\theta) = c_0 + d \ \theta,
\label{def:spiral_polar_equation}
\end{equation}
where $r$ and $\theta$ are the usual polar coordinates. The parameter $c_0$ moves the centerpoint of the spiral outward from the origin, while $d$ controls the distance between the spiral arms.
Actually, in the top-left panel of figure~\ref{fig:r_theta_S_2D_and_3D}, we rather show $r_p \equiv 1 - \sqrt{x^2 + y^2}$ instead of $r$ (for pedagogical reasons only, to focus on a clump rather than a void region), together with $\theta \equiv \frac{1}{\pi} \atantwo(y,x)$ where the $\atantwo$ function generalizes the relation $\theta = \arctan(y/x)$ which holds only for $x>0$. The $1/\pi$ factor simply keeps the field in the normalized range $[-1, 1]$ for convenience.

To construct an actual spiral-shaped scalar field in the plane, we consider $\lambda \equiv r_p - c_0 - d \ \theta$, a local length that measures how far a given point is from the spiral (\ref{def:spiral_polar_equation}). This local length is then given as argument to a suitable filter, for example the smoothed top-hat
\begin{equation}
T(s) \equiv \frac{1}{2} \left(\tanh \frac{s + w/2}{\ell} - \tanh \frac{s - w/2}{\ell} \right),
\label{def:T}
\end{equation}
which is a function such that $T(s)$ equals $1$ in a region of width $w$ near the origin $s = 0$, and equals $0$ elsewhere with a smooth transition from 1 to 0 of thickness controlled by the length $\ell$.
The field $T(\lambda)$ is a field with a spiral shape, because $T$ selects the regions of space where $\lambda$ is close to $0$, up to a certain width $w$. Most importantly, so far $r_p$ is a 2D field, but in the next section we will replace it by a 3D field related to $\vec{R}$ from (\ref{def:R}), turning $T(\lambda)$ into a 3D scalar field with spiral-shaped sheet-like structures. Anticipating this, we refer to $T(\lambda)$ as a sheet.

At this point, we obtain further guidance from the current density field $\vec{j}$ as obtained in actual DNS studies, where it appears relevant to distinguish two types of sheets in the modelling of turbulent magnetized flows. Indeed, they suggest clearly a bimodality in (i) intense (i.e. high $|\vec{j}|$ regions), which are thin, and relatively rare sheets (i.e. intermittent), and (ii) more diffuse weaker $|\vec{j}|$ regions distributed in thicker, and more abundant sheets (i.e. more volume-filling), which surround the intense sheets.

Therefore, we define an intense filter $T_i$ and a diffuse filter $T_d$, which are identical to $T$ in (\ref{def:T}) with differing numerical values for the parameters ($w_i, \ell_i$) and ($w_d, \ell_d$) respectively: $w_i$ and $\ell_i$ are smaller than $w_d$ and $\ell_d$, to mimic the fact that intense sheets are thinner and less blurry than diffuse sheets.
Secondly, as seen in the top-middle panel of figure~\ref{fig:r_theta_S_2D_and_3D}, the $\atantwo$ function introduces a discontinuity where $\theta = \pm 1$. As a simple work-around to avoid jumps in our magnetic field model, we impose a spatial-dependence to the width $w_i$ of intense sheets through the prescription (recall that $\theta \in [-1, 1]$)
\begin{equation}
w_i = w_i^\text{max} \cos(\pi \theta/2),
\label{def:w_i}
\end{equation}
where $w_i^\text{max}$ is a constant. As a result, wherever $\theta$ is discontinuous, intense sheets become infinitely thin, and therefore vanish. Intense sheets are then also less volume-filling, and hence more intermittent, as it appears in DNS simulations.
It turns out to be unnecessary to do the same for the width $w_d$ of diffuse sheets, because these sheets have weak amplitudes, so their discontinuities are smoothed out when taking the Biot-Savart law (\ref{def:B}) in the last step of our construction.
Finally, as illustrated in the top-right panel of figure~\ref{fig:r_theta_S_2D_and_3D}, to model the fact that diffuse sheets are numerous and surround intense sheets, we generalize our filtering to
\begin{equation}
S(\lambda) \equiv T_i(\lambda) + \epsilon \ T_d(\cos(k_d \lambda)).
\label{def:S}
\end{equation}
The first term corresponds to an intense sheet, and the second to several diffuse sheets. Indeed, instead of $T_d(\lambda)$ we consider $T_d(\cos(k_d \lambda))$, which gives rise to as many sheets as there are zeros in $\cos(k_d \lambda)$, i.e. $k_d$ controls the number of diffuse sheets. Moreover, in regions where $\lambda \sim 0$ this cosine does not vanish so that diffuse sheets are absent, which adequately gives room to the intense sheet $T_i(\lambda)$ sitting there. Lastly, the free parameter $\epsilon$, assumed to be small, makes diffuse sheets more diffuse than intense sheets by controlling their relative amplitude.

To sum up, for our 2D field $r_p$ and angle $\theta$ as in the first two panels of figure~\ref{fig:r_theta_S_2D_and_3D}, $S$ given by (\ref{def:S}) is a field of nested, 2D spirals where the central one is intense, as shown in the top-right panel of that same figure. This constitutes the basic structure of an (isolated) eddy in our model. The key point of the next section is that we will insert in (\ref{def:S}) a 3D (random) scalar field instead, such that $S$ will indeed be a field of 3D sheets with artificially constructed spiraling behavior. Note that thus far, our spirals have constant curvature, to be remedied in what follows as well.

\subsection{Randomly distributing eddies}
\label{sec:adding_randomness}

We now present an efficient way (i.e. a simple single step) to simultaneously
(i)~extend from 2D to 3D the above considerations, (ii)~introduce non-trivial spatial variations of the curvature of the sheets, and (iii)~distribute eddies in the whole domain, with the properties of the sheets (size and wiggliness) controlled by a few parameters.

As mentioned in the preliminaries, our fractional Gaussian field $\vec{R}$ given by (\ref{def:R}) is a poor stochastic model for a turbulent magnetic field. In the bottom-left panel of figure~\ref{fig:r_theta_S_2D_and_3D} we show a 2D cut of a realization of its norm, $R \equiv |\vec{R}|$. The 3D scalar field $R$ consists of an ensemble of nearly spherical clumps of various sizes, randomly distributed throughout space. The fact that this field does not resemble actual turbulent structures is related to the (visual appearance) shortcoming we alluded to in our introduction of present multiplicative chaos models for hydro turbulence. The clumpiness of $R$ and the typical size of its largest clumps are readily controlled by the Hurst parameter $h_R$ and the cut-off $L_R$ in (\ref{def:R}), respectively. Having noticed this, we will now use this clumpy field to build spiral-shaped structures swirling around intense clumps.
Hence, we are not going to use $\vec{R}$ as a magnetic field vector $\vec{B}$ model, but as our foundation to build a current $\vec{c}$, to plug in the formula (\ref{def:B}) for~$\vec{B}$.

We now have a natural `radius field' $R$, but in analogy with the construction of 2D spirals, it remains to find a relevant angle $\theta$. This is indeed possible noticing that we may also write $\theta = \atantwo\left( \partial_y r, \partial_x r \right) /\pi$, a relation that becomes clearer after checking that it does reduce to the standard $\arctan(y/x)$ for $x>0$. With this new viewpoint, it is now natural to define, for the 3D case,
\begin{equation}
\theta_R \equiv \frac{1}{\pi} \atantwo\left( \partial_y R, \partial_x R \right).
\label{def:theta_with_R}
\end{equation}
Finally, we redefine the length $\lambda$ as
\begin{equation}
\lambda_R \equiv R - c_0 - d \ \theta_R.
\label{def:lambda_with_R}
\end{equation}
Our motivation for these peculiar definitions is purely geometrical, in the sense that we introduce them independently of the dynamical equations. However, an expression such as (\ref{def:theta_with_R}) should not be surprising, since dot products between fields and gradients (and therefore angles) are omnipresent in (magneto-)fluid dynamics, notably with the advection operator $\vec{v} \cdot \vec{\nabla}$. 
Note that, in this 3D case, we could likewise consider a second angle, inspired from the $\phi$ angle of spherical coordinates, but we deliberately keep our model as elementary as possible.

All in all, our magnetic field model $\vec{B}$ is the modified Biot-Savart law (\ref{def:B}) with the `current' vector field in it taken as
\begin{equation}
\vec{c} \equiv S \vec{R},
\label{def:c}
\end{equation}
i.e. $\vec{c}$ starts from the fractional Gaussian field $\vec{R}$ given by (\ref{def:R}), scaled by a sheet-like field with a spiral structure $S$ given by (\ref{def:S}), where the Top-Hat functions $T_i$ and $T_d$ are given by (\ref{def:T}), the angle $\theta_R$ by (\ref{def:theta_with_R}) and the length $\lambda_R$ by (\ref{def:lambda_with_R}).
We name our model $B\times C$, which stands for `magnetic fields from multiplicative chaos' in reference to notably \cite{Kahane85, ChevillardHDR, DurriveEtAl20}.

We can motivate our construction as follows. Evidently, the core of turbulence studies is to understand and be able to model the intricate interactions between scales in turbulent fields. A classical paradigm is to consider as total field a split into a sum of fields of different nature, e.g. constituted as an ordered (strong background) plus a turbulent field, or an equilibrium plus a perturbed field. An archetypical example is the mean-field dynamo theory where the magnetic and velocity fields are split into large-scale, mean-field parts and small-scale, fluctuating parts \citep{Rincon19}. In this paper, we introduce another procedure when we use the fGf $R$ field. We effectively introduce a scale-splitting linked to the correlation length scale of $R$: inside each `blob' of $R$ (cf bottom-left panel of figure \ref{fig:r_theta_S_2D_and_3D}) a spiral-shaped eddy forms, while on larger scales, beyond $R$'s correlation length, the eddies decorrelate. Since we expect the statistics of our field to become Gaussian on large scales (see also PDFs of increments further shown in figure~\ref{fig:DNS_vs_BxC_PdfsOfIncrements_B_field_and_j_field}), it seems appropriate to use a Gaussian field, such as a fGf.

\subsection{Mimicking a time evolution}

A particular feature of the present type of modeling is that it consists in applying a \textit{deterministic} transformation to a given white noise. Being deterministic, once a realization of the white noise is chosen, we may transform the magnetic field smoothly by varying continuously the parameters ($L, h, \eta, \epsilon, \dots$). This can be used to emulate a(n artificial) time evolution: to each parameter $p$ we give a simple time dependence $p = \bar{p} + \sigma_p \sin (\omega_p t + \phi_p)$, i.e. the value of $p$ oscillates around a mean value $\bar{p}$, with an amplitude $\sigma_p$, at a frequency $\omega_p$, and a phase shift $\phi_p$. It is paramount to choose different phase shifts for the various parameters. The oscillations will then be out-of-phase, which avoids spurious periodicities. In other words, we thus move continuously in a rather chaotic way into the parameter space. An animation exemplifying this can be found at \footnote{See Supplemental Material at [URL will be inserted by publisher] for our animation.} for a $512^3$ resolution.

\section{Comparing our model to a DNS}
\label{sec:DNS_vs_BxC}

In this section we analyze a realization of a magnetic field built with our $B\times C$ model, as well as its corresponding current density field, and we compare them to a realization of a magnetic field and current density generated using a DNS, to assess the realism of our model.

Note from the outset that we expect our model to be primarily useful (i) to generate extremely high resolution fields (including a mock time evolution) that are out of reach of DNSs, and (ii) to reduce drastically the time needed to create non-trivial initial conditions for DNSs. We therefore will assess whether our model can reproduce with much reduced resources various aspects also present in a given DNS. It is to this end that we ran a full DNS. The latter will constitute some reference data, considered as `realistic', and in this part of the paper we show by means of a series of side-by-side comparisons, that our model shares many properties of this DNS, both qualitatively (notably sheet-like structures with appealing visual aspects) and quantitatively (notably providing evidence of intermittency, and the expected shape for power spectra, namely a well-defined power-law behaving inertial range between clear large-scale and small-scale cut-offs).
Naturally, since our model is a fast parametric model, future work could easily extend it with an automated systematic parameter survey, such as Monte Carlo Markov Chain analyses.

This part of the paper is organized as follows. We give details on how we implemented numerically our DNS and our model magnetic fields. We carry on by comparing the DNS and $B \times C$ fields in five ways. First we compare the resources required to generate them, then we inspect their visual aspects (2D slices as well as 3D appearance, with both scalar and vector visualizations), after which we provide several quantitative comparisons using the standard statistical tools of turbulence studies, namely power spectra, PDFs of increments for $B$ and $j$, structure functions and spectra of exponents, supplemented with a Partial Variance of Increments analysis.

\subsection{Numerical implementation}
\label{sec:numerical_implementation}

Throughout the paper, unless otherwise stated, the fields have a resolution corresponding to $N^3=1024^3$ collocation points.

The DNS dataset considered is a snapshot at the temporal peak of total dissipation from a pseudo-spectral simulation of decaying 3D isotropic MHD turbulence that was performed with the ALIAKMON code \citep{Momferatos15}. The non-linear terms in the equation were de-aliased using the standard two-thirds rule, while advancement in time was performed by a fourth-order Runge-Kutta method. The product of the maximum wave-number that was represented in the simulation with the Kolmogorov micro-scale was at all times kept above 2. At the temporal peak of total dissipation, the Taylor micro-scale Reynolds number is approximately equal to 270, while the Reynolds number based on the integral length scale is approximately equal to 2000. The initial condition used is a superposition of a large-scale Arnol'd-Beltrami-Childress (ABC) flow at $|k| = 2$ and a Gaussian random field with an exponentially-decaying energy spectrum.

For $B \times C$, we detail the reasoning that lead us to the chosen numerical values of the parameters. Note that in our code the box size is taken equal to unity, so the numerical values of the lengths below should be read as percentages of the box size.

First we chose the parameters for the fGf $R$, because the fGf directly controls the typical size of the large eddies, as illustrated by the bottom panels of figure~\ref{fig:r_theta_S_2D_and_3D}. In order to obtain about ten large eddies along each direction of our data cube, we took $L_R$ equal to about a tenth of the box size, specifically $L_R = 0.075$. Then, for the inertial range to be as large as possible, we needed to choose $\eta_R$ as small as possible, but as previously mentioned, at the same time this parameter should be large enough for the gradients of this random field to be well approximated. A usual trade-off in such models \citep[e.g.][]{PereiraEtAl16, DurriveEtAl20} is to take $\eta_R = 3 dx$, where $dx = 1/N$ is the pixel size on our grid of size $N=1024$. Finally, the Hurst parameter $h_R$ of the fGf controls how smooth $R$ is, and consequently, given the construction, it controls how wiggly the sheets are (cf. bottom panels of figure~\ref{fig:r_theta_S_2D_and_3D}). Since in our reference DNS data the sheets are particularly smooth, sometimes even almost flat, we were led to choose a very small numerical value for this Hurst parameter, and took $h_R = 0.05$.

For the Biot-Savart law (\ref{def:B}) we took $L = 0.3$ to integrate on sufficiently large regions for the magnetic field to span on large scales, as in our DNS. The choice $h = 2$ was based on enabling the magnetic field to have enough power at small scales, since Hurst parameters control the slope of the power spectrum, and the standard value $\eta = 3/N$ was chosen as for $\eta_R$ above.

For the properties of the sheets, we focused on the parameters controlling the spiral shapes.
Given the properties of Archimedean spirals, we chose $c_0=0.3$ to offset the centerpoint of the spirals from the origin to avoid having sheets converging artificially at the same points, and $d=0.2$ for the sheets to be well separated, as in the DNS.
Otherwise, in order to make our intense sheets very thin, as in our reference DNS, we chose a width several orders of magnitude smaller than the box size, namely $w_i^{\mathrm{max}}=3\mathrm{e}{-5}$ and $\ell_i=5\mathrm{e}{-3}$.

We then constructed the diffuse sheets relatively to the intense ones: In our reference data diffuse sheets appear typically an order of magnitude wider, hence $w_d=0.05$, and being `diffuse' translates into $\ell_d=0.2$ to be an order of magnitude larger than in intense sheets (the filter thus being far less steep). In addition, from (\ref{def:S}) it is clear that $k_d$ controls the number of sheets swirling inside a given eddy, measured in multiples of $2 \pi$. The choice $k_d=6 \pi$ leads to a few diffuse sheets and an appropriate volume-filling aspect. To make sheets diffuse and hence less intense, it is natural to weigh their amplitude relative to the intense sheets by a number of the order of a percent, hence $\epsilon=5\mathrm{e}{-3}$.

Finally, in Biot-Savart's law we smoothed the truncation of the integration region to a ball of radius $L$, by multiplying the kernel by $0.5(1-\tanh{(r-L-\ell_c)/\ell_c})$ with $\ell_c = 0.23 L$. Second, noticing that in places where $|\vec{\nabla} R| \sim 0$, such as in the origin of the plane in the top row of figure~\ref{fig:r_theta_S_2D_and_3D}, many sheets converge in a relatively artificial manner, we multiplied $S$ by $1-\exp(-|\vec{\nabla} R|^2/10)$, and found that this improved slightly the results.

\subsection{Comparison 1: Required resources}
\label{sec:comparing_required_resources}

\begin{figure}
\includegraphics[scale=0.6]{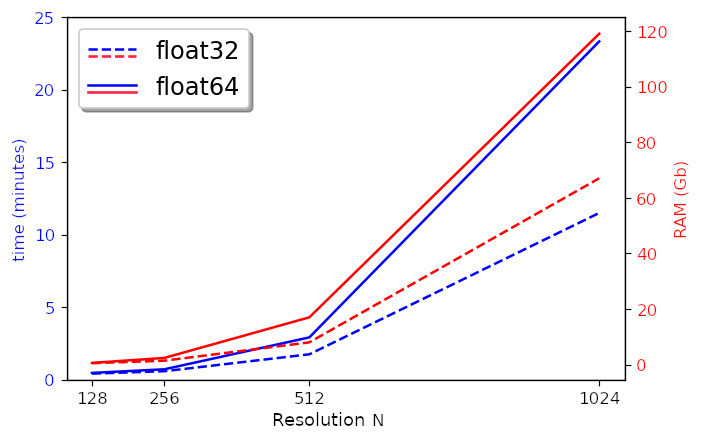} 
\caption{\label{fig:Scalings} (Color online) Resources required to generate a magnetic field realization with our $B \times C$ code (written in Python). Computing time as a function of resolution is plotted in blue (left $y$-axis), and the required RAM memory in red (right $y$-axis), performed with a 40-logical-cores desktop. Continuous lines are for double precision (float64) calculations, and dashed lines for single precision (float32). Hence, $1024^3$ data are generated in about 10 to 25 minutes, depending on the precision needed.}
\end{figure}

The important difference between $B\times C$ and our DNS run is the resources used: the reference DNS required about 50 000 core hours (on an HPC system with 8-core Intel E5-2670 Xeon processors running at 2.60 Ghz) while a magnetic field with our code is generated in less than half an hour on a desktop with 40 logical cores, as detailed in figure~\ref{fig:Scalings}. Our model is very fast because (\ref{def:B}) and (\ref{def:R}) are nothing but convolution products, i.e. simple products in Fourier space. In contrast to the original HD and MHD models of this kind, where much more intricate nonlinearities were used to mimic turbulence statistics, this aspect makes our geometric, parametrized construction scalable to extreme resolutions, beyond those achievable by DNS on modern supercomputers, and only bound by local memory requirements. Order $500^3$ realizations are feasible on any laptop, while modern desktops can easily generate far larger fields.

\subsection{Comparison 2: Visual aspects}

\begin{figure*}
\includegraphics[scale=0.33]{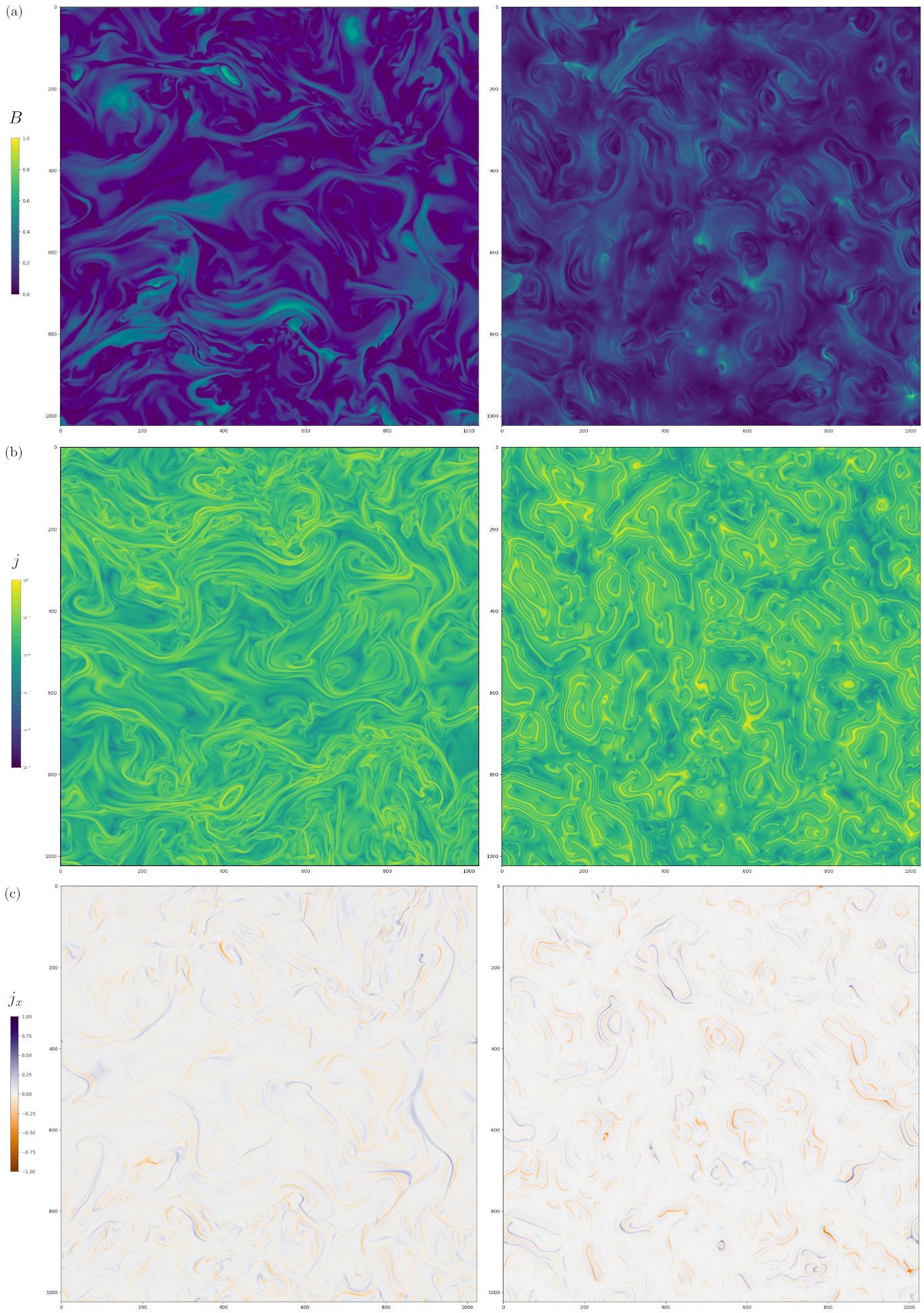} 
\caption{\label{fig:DNS_vs_BxC_normB_lognormj_jx} (Color online) Figures~\ref{fig:DNS_vs_BxC_normB_lognormj_jx}, \ref{fig:DNS_vs_BxC_contours_0d10_0d30_0d60} and \ref{fig:DNS_vs_BxC_VectorSheets} are visual comparisons of fields generated with a DNS (left column) to fields generated with our $B\times C$ model (right column). In the present figure: (a) in the top row are slices of $B$, the norm of the magnetic field, (b) in the middle row are slices of $j$ in logarithmic scale, the norm of the current density, and (c) in the bottom row are slices of $j_x$, the $x$-component of the current density, which shows some vector information (orientation of $\vec{j}$). From figures~\ref{fig:DNS_vs_BxC_normB_lognormj_jx}, \ref{fig:DNS_vs_BxC_contours_0d10_0d30_0d60} and \ref{fig:DNS_vs_BxC_VectorSheets} we conclude that, while $B\times C$ fields are generated using several orders of magnitude less resources, they have a similar visual aspect than the DNS.}
\end{figure*}

\begin{figure*}
\includegraphics[scale=0.5]{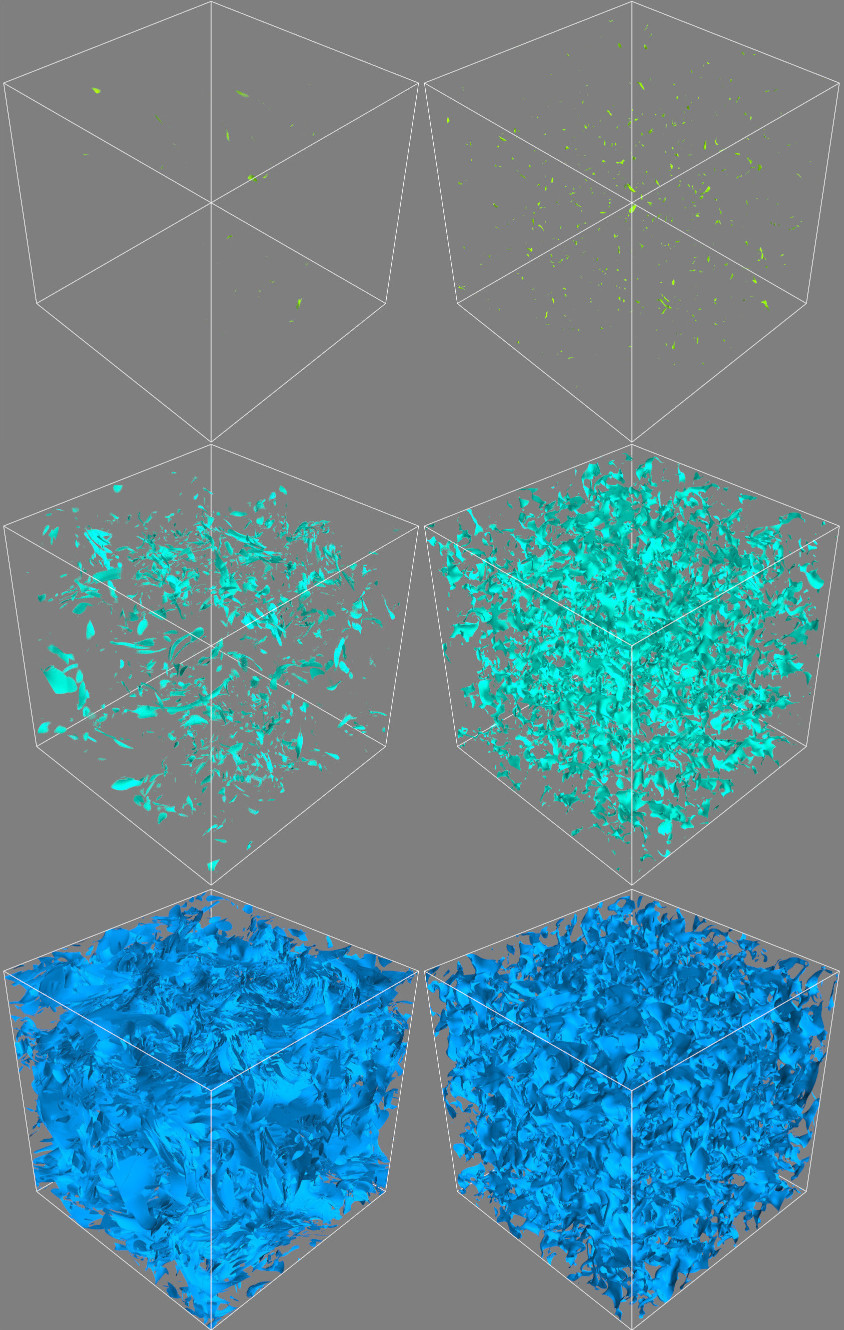} 
\caption{\label{fig:DNS_vs_BxC_contours_0d10_0d30_0d60} (Color online) Continuation of figure~\ref{fig:DNS_vs_BxC_normB_lognormj_jx}. Iso-contours of $j$, the norm of the current density, are shown for values of $60\%$ (top row), $30\%$ (middle row) and $10\%$ (bottom row) of the maximal value. The volume-filling and the shape of the structures of the $B\times C$ field at different amplitudes of $j$ matches qualitatively that of the DNS.}
\end{figure*}

\begin{figure*}
\includegraphics[scale=0.55]{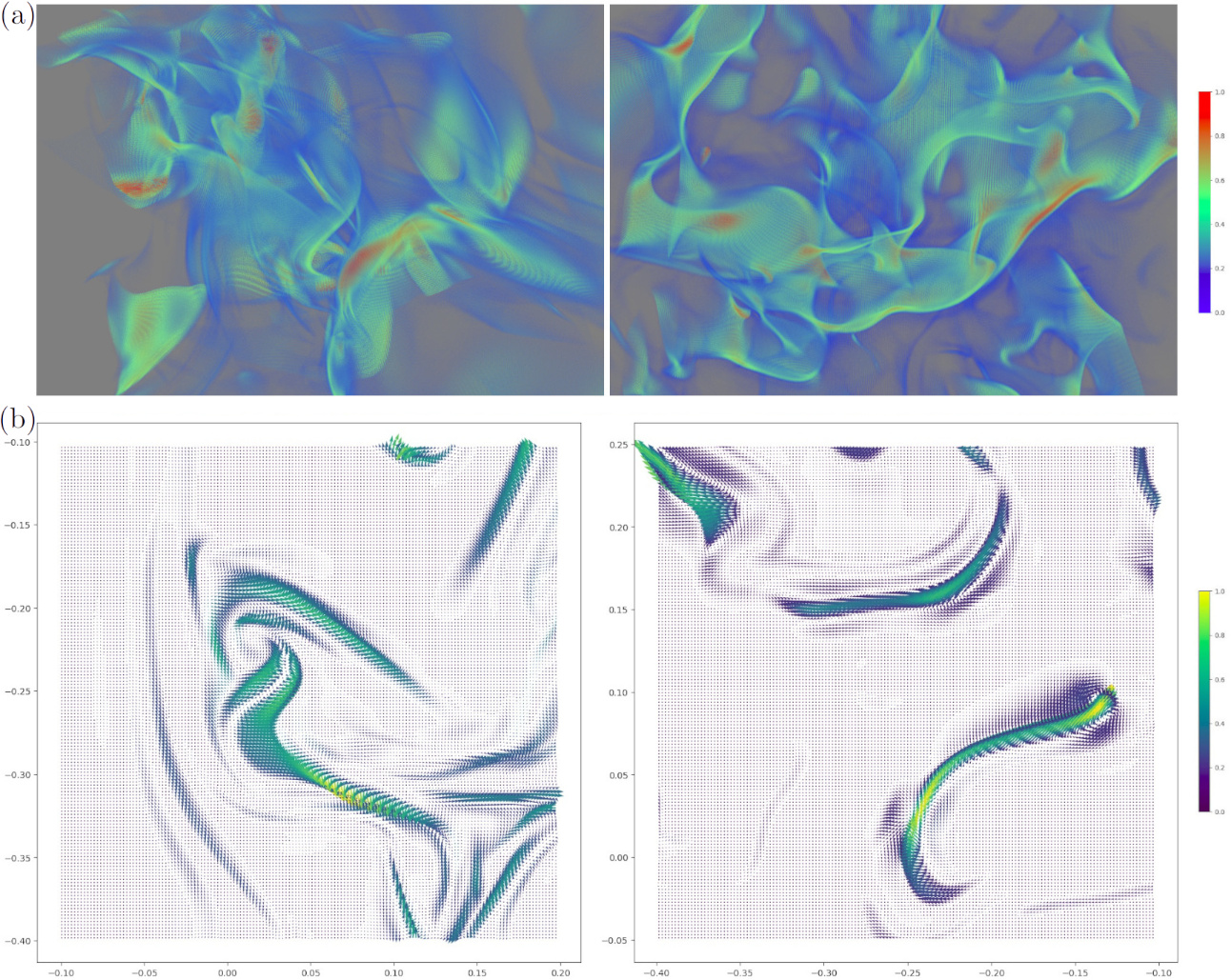} 
\caption{\label{fig:DNS_vs_BxC_VectorSheets} (Color online) Continuation of figures~\ref{fig:DNS_vs_BxC_normB_lognormj_jx} and \ref{fig:DNS_vs_BxC_contours_0d10_0d30_0d60}. (a) In the top row are 3D vector visualizations and (b) in the bottom row are 2D vector cuts. While figures~\ref{fig:DNS_vs_BxC_normB_lognormj_jx} and \ref{fig:DNS_vs_BxC_contours_0d10_0d30_0d60} show the distribution of the sheets throughout the whole space, here we have zoomed on specific regions to reveal finer details of some clusters of sheet-like structures.}
\end{figure*}

In figures~\ref{fig:DNS_vs_BxC_normB_lognormj_jx}, \ref{fig:DNS_vs_BxC_contours_0d10_0d30_0d60} and \ref{fig:DNS_vs_BxC_VectorSheets}, the left columns correspond to the DNS and the right columns to our $B\times C$ model.
The first row of figure~\ref{fig:DNS_vs_BxC_normB_lognormj_jx} shows the magnetic fields, while all the other figures correspond to the current density fields, which $B \times C$ aims at reproducing. We insist that $\vec{j}$ here is computed, as it should, by taking the curl of the magnetic field (\ref{def:B}): it does \textit{not} simply correspond to $\vec{c}$ given by (\ref{def:c}), because (\ref{def:B}) is a modified Biot-Savart formula.

In figure~\ref{fig:DNS_vs_BxC_normB_lognormj_jx} we start by exhibiting 2D slices of the norms of $\vec{B}$ and $\vec{j}$, in the top and middle rows respectively. It appears that the $B\times C$ fields have a fluid aspect in the sense that smooth variations alternate sudden concentrated structures. The overall size distribution of larger and smaller patches, in both the magnetic field magnitude variation and in the current intensity, is fairly similar between DNS and our model.
An advantage of this construction is that the properties of the sheets are easily controlled by a few parameters: $c_0$ and $d$ in $\lambda$ given by (\ref{def:lambda_with_R}) control the (deterministic) shape of individual spirals constituting the sheets, while $L_R$ and $h_R$ in the fGf $R$ given by (\ref{def:R}) control respectively the typical size of the swirling regions and how wiggly the sheets are. For example in the $B\times C$ realization shown here, we deliberately chose a very small numerical value for the Hurst parameter $h_R$. This way $R$ is very smooth (cf. bottom-left panel of figure~\ref{fig:r_theta_S_2D_and_3D}) so that the sheets are not very wiggly, as we observed in our reference DNS data.

Then, in the bottom row of figure~\ref{fig:DNS_vs_BxC_normB_lognormj_jx}, to provide some vector information, we show a 2D slice (the same as for the two rows above) of $j_x$, the $x$-component of $\vec{j}$. In the blue regions $j_x$ is positive, while it is negative in the red regions. Comparing the DNS and our analytic reproduction (the left and right columns) we conclude that $B\times C$ reproduces, qualitatively and statistically speaking, the orientation of the sheets. This is an extremely important finding, since we targeted this 3D turbulent vector correspondence from the outset, not just a scalar reproduction.

In figure~\ref{fig:DNS_vs_BxC_contours_0d10_0d30_0d60} we show iso-contours of $j$ at $60\%$, $30\%$, and $10\%$ of its maximal value, in the top, middle and bottom rows respectively. These 3D visualizations confirm that the $B\times C$ current density field is indeed composed of 3D sheets with non-trivial shapes (non-uniform curvature and wiggly edges). The distribution (i.e. the positions, the orientations and the volume-filling aspect) of the sheets is rather realistic, in the sense that intense $j$ regions are not volume-filling, which is one known facet of MHD intermittency.

In figure~\ref{fig:DNS_vs_BxC_VectorSheets} we show yet more vector information, complementing the bottom panel of figure~\ref{fig:DNS_vs_BxC_normB_lognormj_jx}. The top row is a zoom into a 3D vector visualization, while the bottom row is a zoom on a 2D vector visualization, both displaying regions with many sheets. We again conclude that the look and feel of $B\times C$ is convincing, and it should be noted that we have not yet attempted to optimize the free parameters involved in any way. This can probably be done in follow-up work, but it is to be stressed that we can easily generate many realisations within hours on desktop resources, which in principle are equally likely, just by changing our starting Gaussian noise model.

\subsection{Comparison 3: Power spectra}
\label{sec:power_spectra)}

\begin{figure}
\includegraphics[scale=0.30]{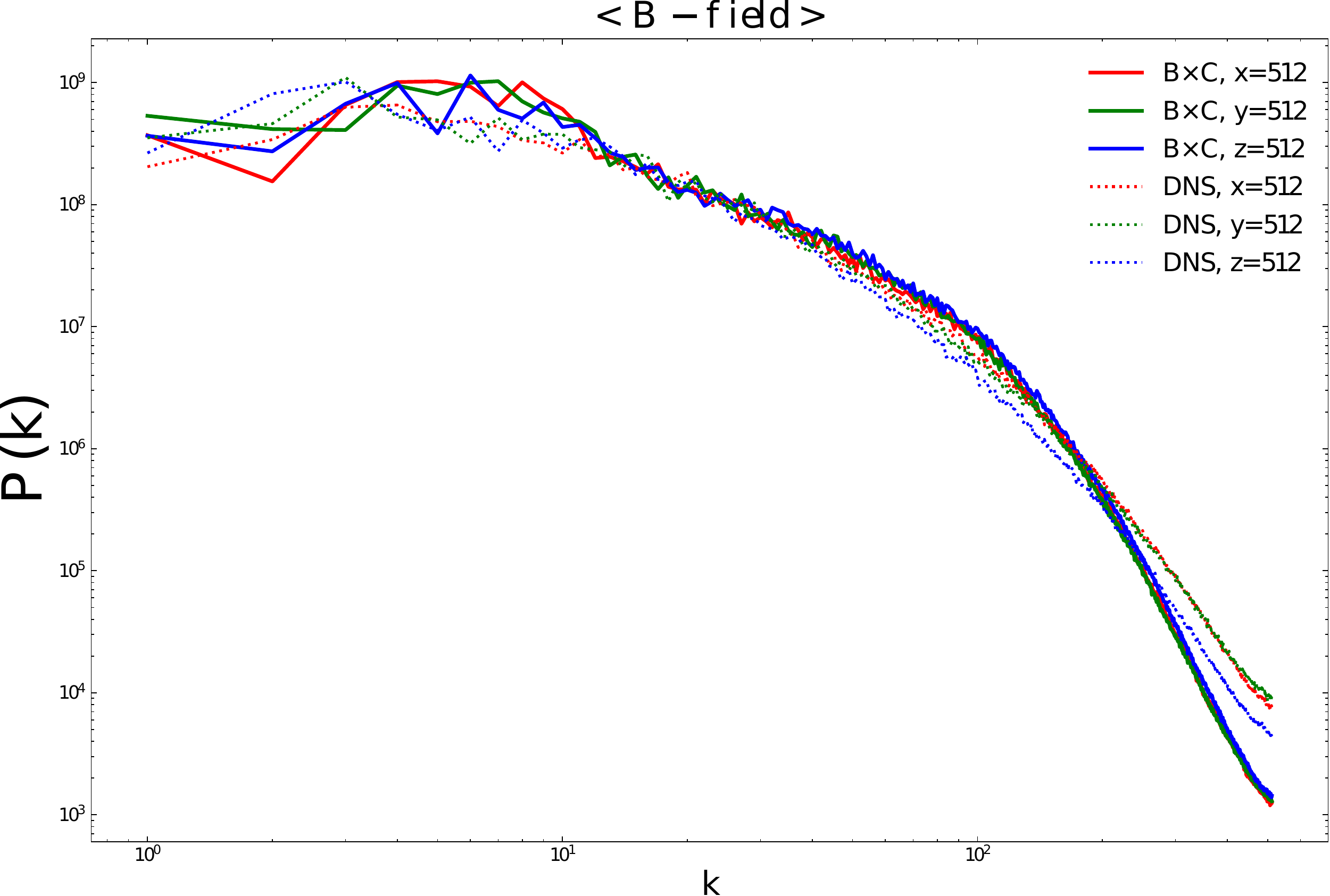}
\includegraphics[scale=0.30]{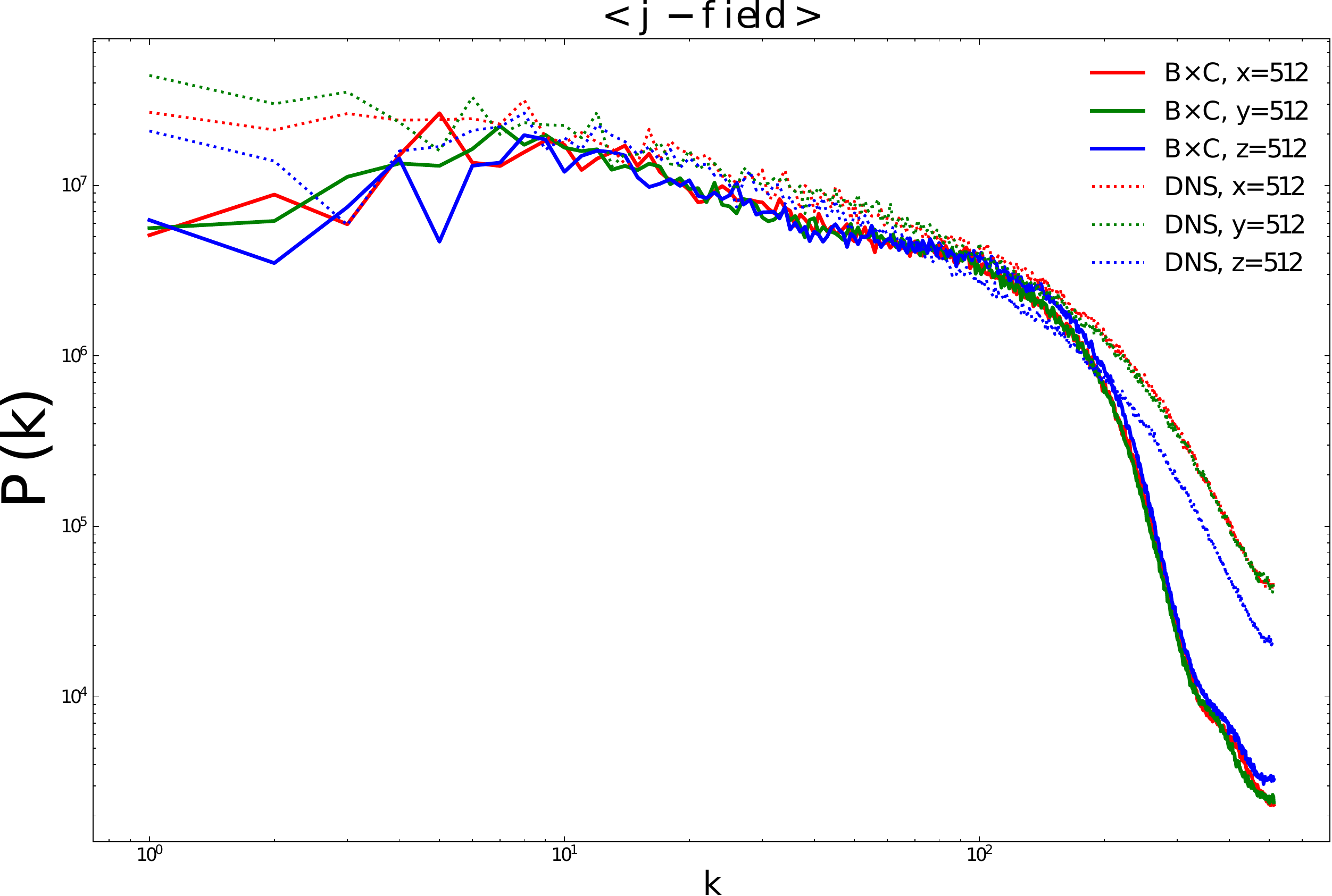}
\caption{\label{fig:powerSpectra_verticalLayout} (Color online) Statistical comparison of the DNS and $B\times C$ fields. 
In the top panel we plot the power spectrum $P(k)$ of the norm of the magnetic field, from three orthogonal 2D slices passing through the center of the data cube. Continuous lines correspond to $B\times C$ and dotted lines to the DNS, where in red the data used is from the slice with fixed $x = 512$ (the resolution being $N = 1024$), in green with fixed $y = 512$ and in blue with fixed $z = 512$. Red, green and blue curves of a given dataset match because the fields are statistically isotropic.
The bottom panel is the same with the current density field. The important point is that the power spectra of the $B\times C$ fields have the characteristic shape of turbulent fields, with a clear power-law inertial range delimited by a large-scale cut-off at small~$k$ and a small-scale cut-off at large~$k$.}
\end{figure}

\begin{figure}
\includegraphics[scale=0.30]{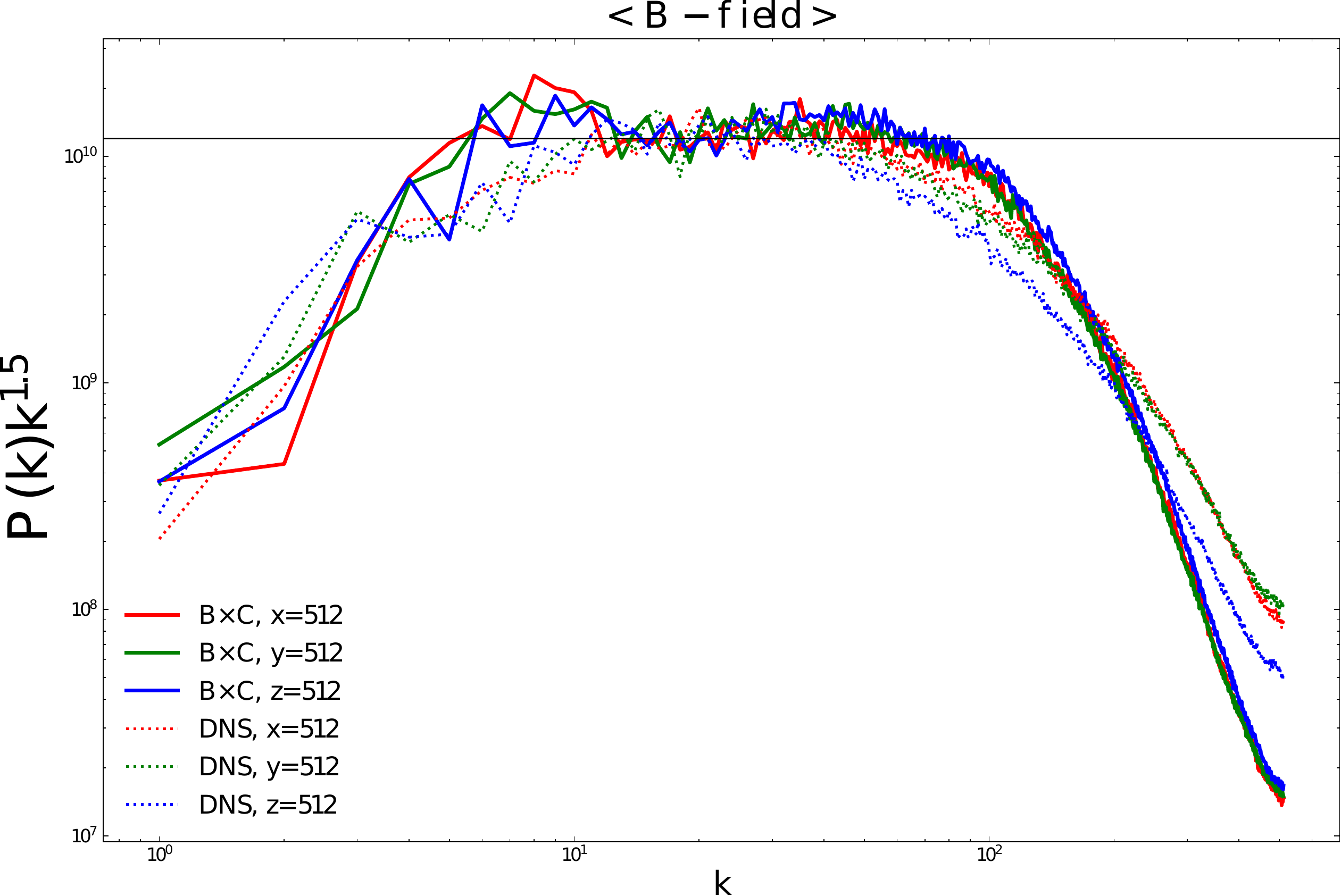}
\includegraphics[scale=0.30]{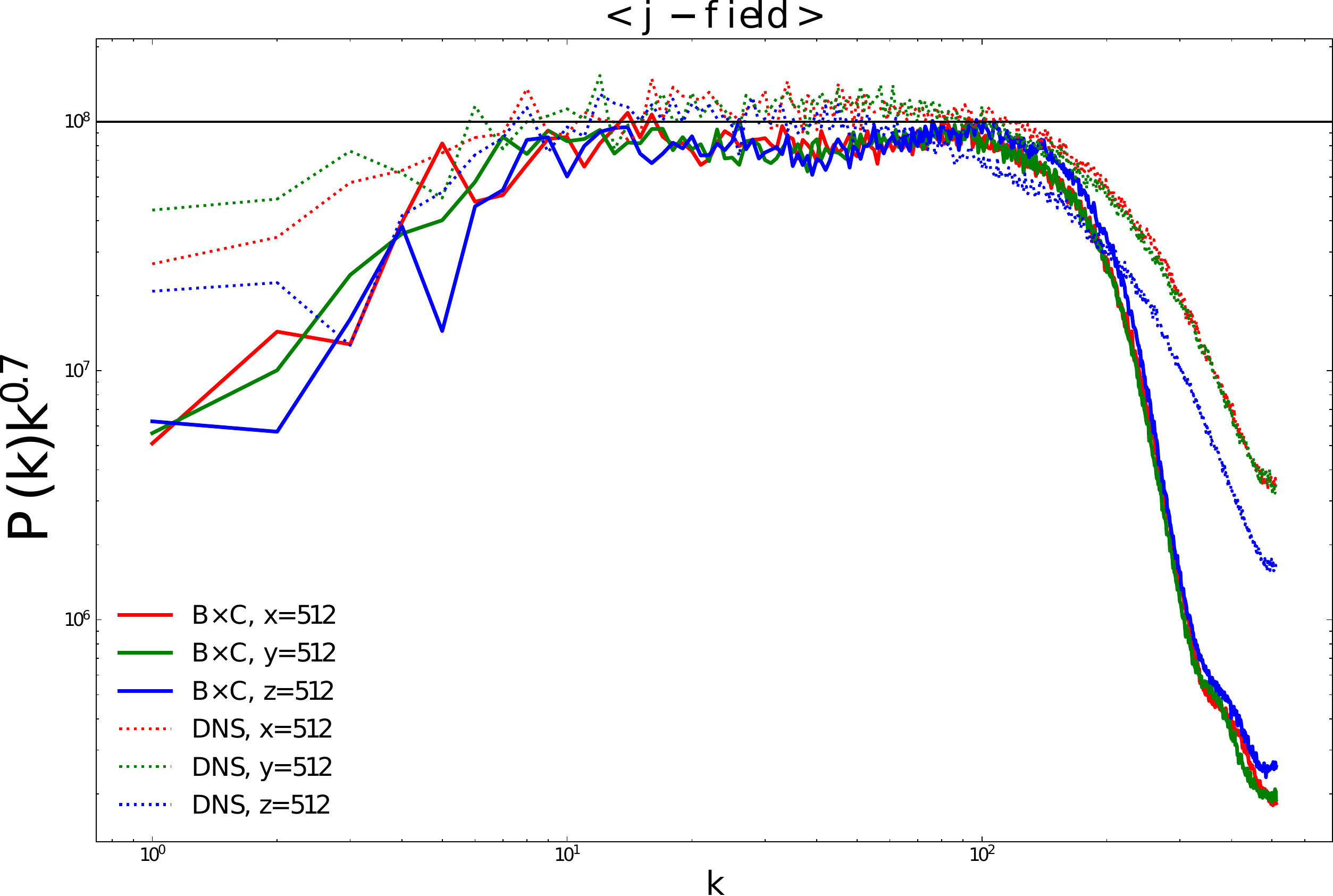}
\caption{\label{fig:powerSpectra_compensated} (Color online) Same as figure~\ref{fig:powerSpectra_verticalLayout}, but where the spectra have been compensated.}
\end{figure}

\begin{figure}
\includegraphics[scale=0.24]{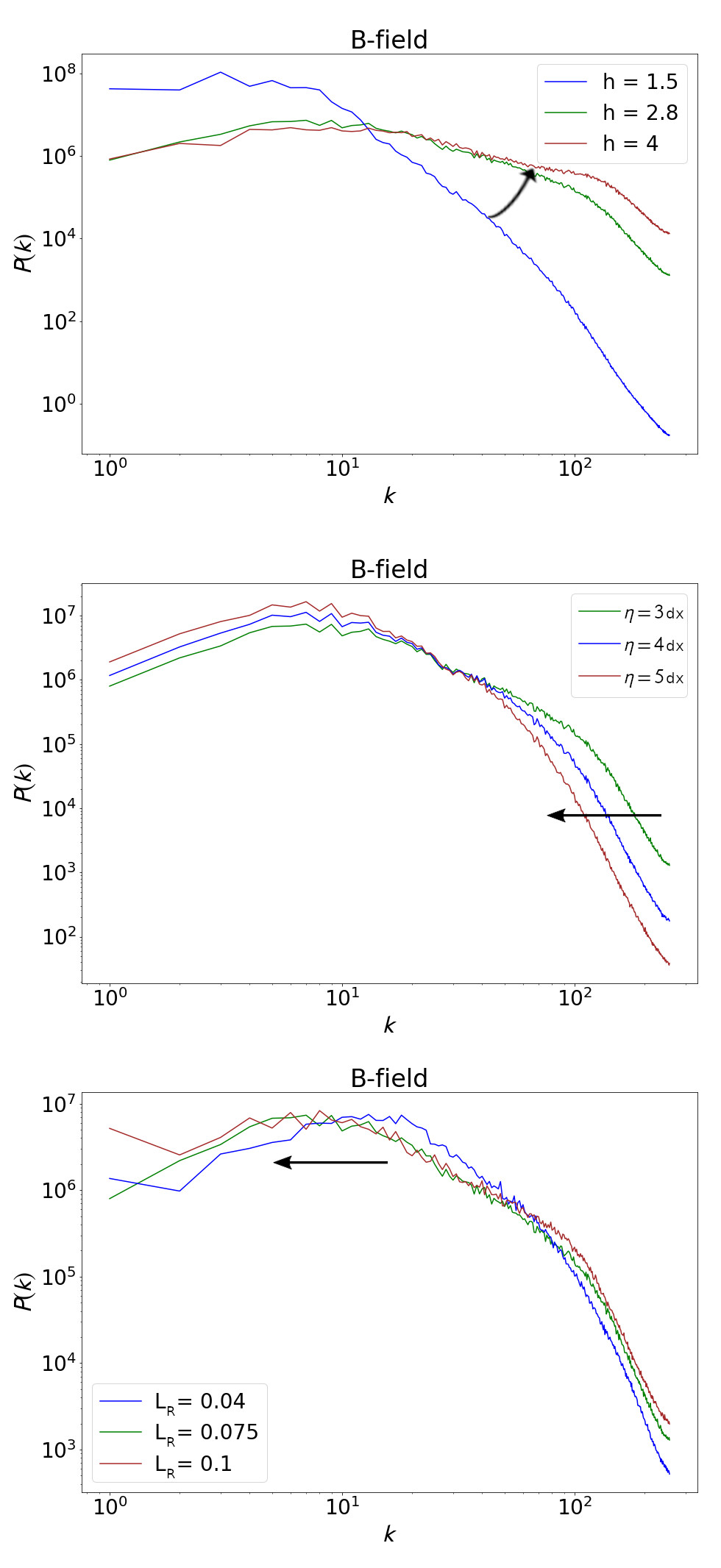}
\caption{\label{fig:Varying_parameters} (Color online) Three illustrations of how the power spectrum of the norm of a $B \times C$-generated magnetic field changes when varying some parameters of the model (namely $h$, $\eta$ and $L_R$ from top to bottom panels), while keeping the other parameters to their reference values. These examples were made at a resolution $N = 512$, and, as in the rest of the paper, the parameter $L_R$ is measured in box-size units and $dx = 1/N$. The black arrows suggest how tweaking these parameters may help fitting a given power spectrum.
}
\end{figure}

Finally, in figure~\ref{fig:powerSpectra_verticalLayout} we computed detailed statistical information to be more quantitative. In the DNS and $B \times C$ code simulations, we consider the power spectrum as a tool to quantify and compare the statistics of the scale dependence of the fluctuations. The power spectrum (P) is defined as the change in kinetic energy (E) as a function of wavenumber (k), $P(k) = dE/dk$. From the isotropic incompressible 3D data of the field, we generate a 1D radially averaged power spectrum \citep{2007A&A...469..595M} from 2D slices along coordinate directions. The 2D field $f(x,y)$ is Fourier transformed, yielding the 2D power spectrum from the amplitude defined as $P(k_x , k_y) = |\Tilde{f}(k_x,k_y)|^2$ where $\Tilde{f}$ denotes the Fourier transform of the field. The collapsed 1D radial average of $P(k_x, k_y)$ between $k$ and $k + dk$, where $k = \sqrt{k_x^2 + k_y^2}$, yields the power spectrum $P(k)dk$. This is shown in Fig.~\ref{fig:powerSpectra_verticalLayout}.
The fact that the red, the green and the blue curves of a given dataset -- which differ in their slice orientation -- overlap each other \footnote{A slight anisotropy at small-scales seems to appear in the DNS spectra. However, we have checked that when sampling our DNS data cube along each direction (x,y, and z) into eight equally-spaced slices rather than only one, the x, y, and z spectra do overlap much closer than on Fig.~\ref{fig:powerSpectra_verticalLayout} as they should, because we thus improve our effective ensemble averaging (assuming the slices are representative of independant realizations) when computing the power spectra.}, stems from the statistical isotropy of the fields. This behaviour would obviously change if we were to combine a $B\times C$ prescription with background guide fields, or given spatio-temporally varying, smooth background magnetic field models.
This figure shows that the $B\times C$ fields, in particular the magnitudes $B$ and $j$, have similar power spectra than that of the DNS. Indeed, they provide a clear proof of concept, and produce fields with power spectra that have the characteristic shape of turbulent fields, namely a large-scale cut-off at small $k$ corresponding to the injection scale, a power-law inertial range at intermediate $k$ which, physically speaking would correspond to the energy cascade, and a clear small-scale cut-off at large $k$ which mimics the effects of dissipation.

Note that we can always scale the magnetic field strength in the $B\times C$ to match the DNS power spectrum at a specific lengthscale, notably with $N_B$. The comparison between power spectra in Fig.~\ref{fig:powerSpectra_verticalLayout} is repeated in compensated form in Fig.~\ref{fig:powerSpectra_compensated}, showing a very acceptable level of agreement, given that no parameter optimization has been performed. The numerical values of the powers in $k$ for the compensations were chosen such that the regression lines of the inertial ranges fit become horizontal.

In figure~\ref{fig:Varying_parameters} we provide three examples of how the power spectra vary when varying the values of the parameters of our model, and how sensitive they are to such variations. Specifically, in the top, middle, and bottom panels we vary respectively $h$, $\eta$ and $L_R$ while keeping all the other parameters to their values of the reference run. We indicate with black arrows how varying these parameters may help tweak a given power spectrum: The Hurst parameter $h$ is a convenient degree of freedom to modify the slope of the spectrum, while $\eta$ and $L_R$ enable refining the cut-offs at the small and large scales respectively. Figure~\ref{fig:Varying_parameters} simply illustrates there are enough degrees of freedom in our model to fit DNS data rather precisely, but this possible optimization is out of the scope of this paper.

\subsection{Comparison 4: PDFs of increments, structure functions and spectrum of exponents}

As turbulent fields are in general not Gaussian fields, power spectra cannot fully characterize a turbulent state. Hence, we now supplement our analysis with  the most common tools of diagnosis in turbulence studies which reveal the existence of intermittent corrections to the scaling of the increments of the fields and their moments (structure functions and spectrum of exponents) with respect to length scale.
Specifically, let us define the increment over a lag $\vec{\ell}$ of the norm $B$ of the magnetic field as the quantity
\begin{equation}
\delta_{\vec{\ell}} B(\vec{x}) \equiv B(\vec{x}+\vec{\ell}) - B(\vec{x}).
\label{def:increment}
\end{equation}
In the following we will also consider the norm $j$ of the current density field, and consider the same expression, replacing $B$ by $j$.

\begin{figure*}
\includegraphics[scale=0.26]{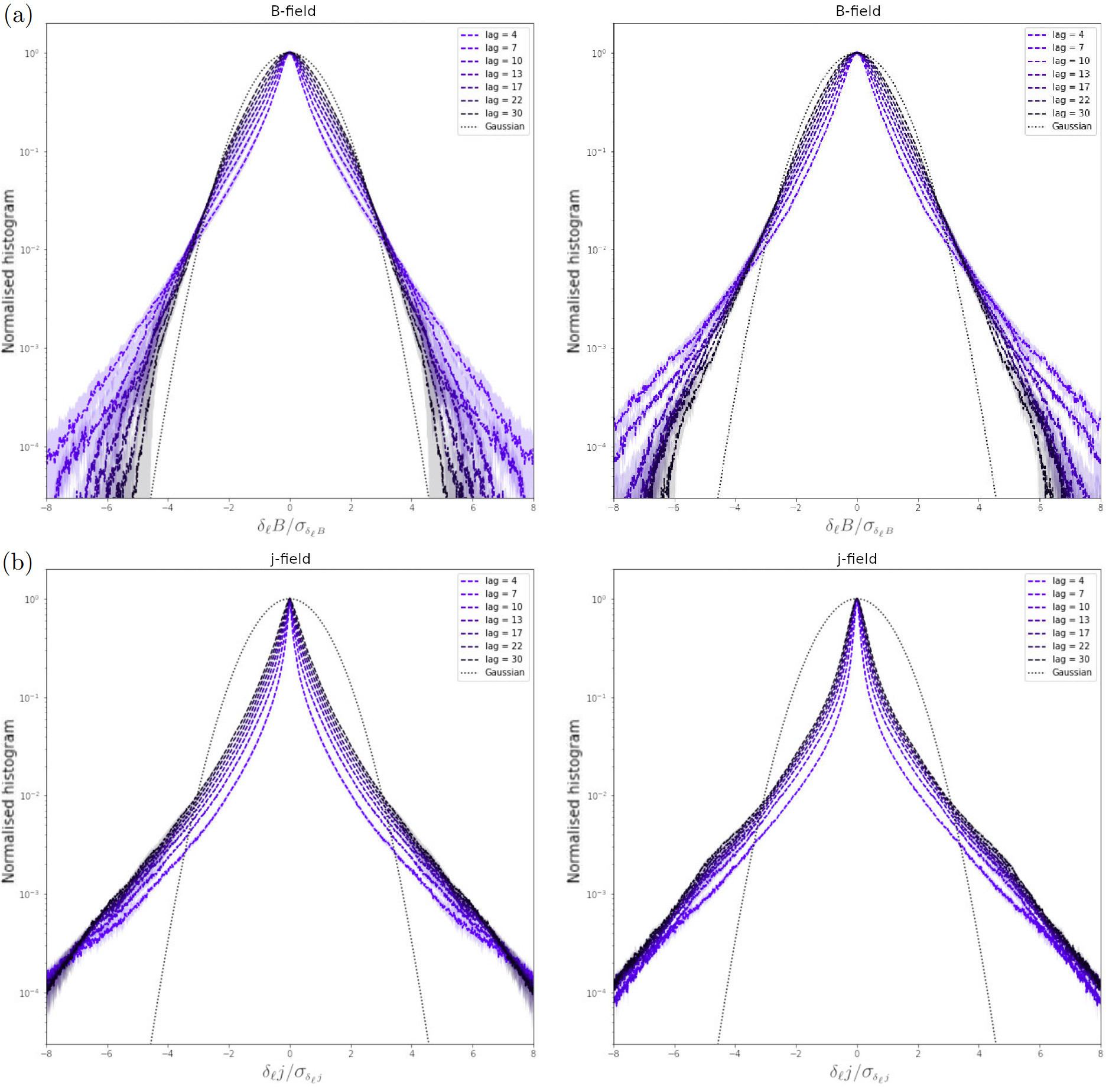} 
\caption{\label{fig:DNS_vs_BxC_PdfsOfIncrements_B_field_and_j_field} (Color online) (a) Top row: PDFs of increments of the norm of the magnetic field generated with our DNS (left column) and that generated with our $B \times C$ model (right column), at lags $\ell = 4, 7, 10, 13, 17, 22, 30$. The dotted black curves correspond to unit-variance Gaussian PDFs. As the lag decreases, the curves deviate from Gaussianity, which is characteristic of intermittency. (b) Bottom row: Same plots using the norm of the current density field instead of $B$. The fact that the plots on the left and right columns look like each other indicates that $B \times C$ generated fields have rather realistic statistical properties.
}
\end{figure*}

A first traditional way to reveal intermittency is to compare the probability density functions (PDFs) of the increments of the considered field to those of a Gaussian field. Indeed, the PDFs of increments in intermittent fields undergo a continuous deformation as the norm $\ell$ of the lag is decreased, the PDF having an almost Gaussian shape at large lags but large tails at small lags. This behavior is a typical signature of intermittency, and the large tails are often called `non-Gaussian wings'. Now, as we saw in the previous section, our fields (both DNS and BxC data) are statistically isotropic since the power spectra of the three slices (cuts along $x$, $y$ and $z$ directions) are very close to one another. Therefore only the norm $\ell$ of the lags matters, and here we compute the PDFs of the increments for $\ell = 4, 7, 10, 13, 17, 22, 30$. For larger $\ell$s the PDFs are near Gaussians. In addition, we use this isotropy to improve our statistics as follows. In practice we compute the PDFs for each direction considering the $x$, $y$ and $z$ slices as independent realizations of a single process, and we show in figure~\ref{fig:DNS_vs_BxC_PdfsOfIncrements_B_field_and_j_field} the median PDF at each lag, with the gray areas indicating the standard deviation from this median.
In this figure, the top row corresponds to PDFs (normalized to unit variance) of $B$ and the bottom row to PDFs of $j$, the left and right columns corresponding to the DNS and BxC data, respectively. In all those plots the departure from Gaussianity is evident, with the aforementionned characteristic continuous deformation when varying the lag. Comparing the two columns, it is manifest that the BxC data does reproduce well the statistics of the reference DNS.

\begin{figure*}
\includegraphics[scale=0.4]{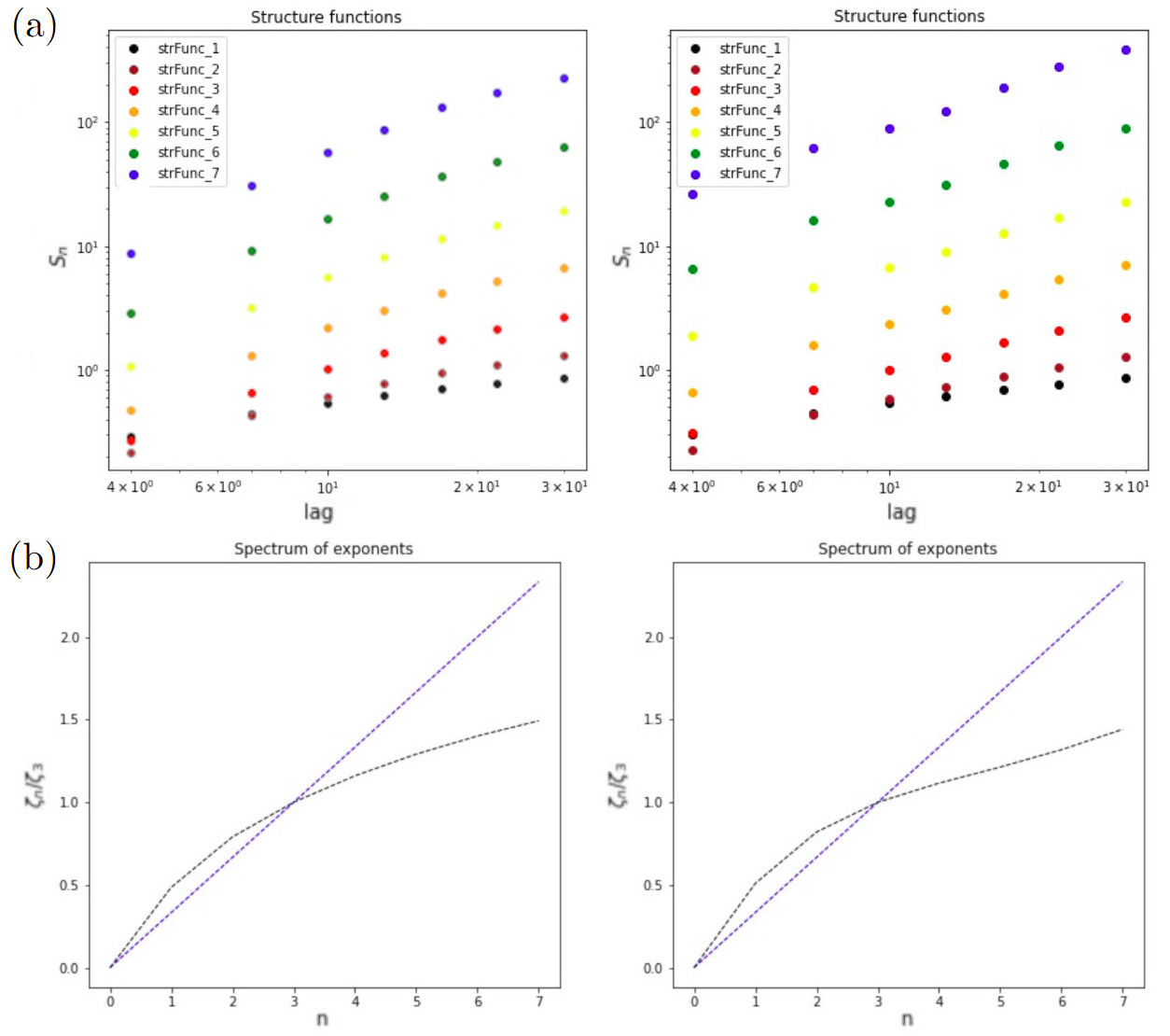} 
\caption{\label{fig:DNS_vs_BxC_StructureFunctions_and_SpectrumOfExponents} (Color online) (a) Top row: The first seven structure functions $S_n(\ell)$ of the norm of the magnetic field generated with our DNS (left column) and that generated with our $B \times C$ model (right column), obtained using the PDFs from figure~\ref{fig:DNS_vs_BxC_PdfsOfIncrements_B_field_and_j_field}. (b) Bottom row: Spectra of exponents resulting from fitting the power-law behaviors of the structure functions shown in the top row. The dashed blue lines correspond to the spectrum of exponents of a non-intermittent field (Kolmogorov scaling), while the dashed black curves correspond to the DNS and $B \times C$ data on the left and right panels, respectively. The pronounced departures of the black curves from the linear laws manifest intermittency, and the similarity between the plots of both columns highlights $B \times C$'s ability to mimic DNS data.
}
\end{figure*}

A second usual way to identify intermittency in isotropic turbulence studies is to analyze the so-called structure functions, and reveal their power-law behavior with respect to scale. Let us define the $n^\text{th}$ order structure function as the $n^\text{th}$ moment of the absolute value of magnetic field increments, namely
\begin{equation}
S_n(\ell) \equiv \langle |\delta_\ell B|^n \rangle,
\label{def:Sn}
\end{equation}
where brackets $\langle \rangle$ denote the expectation value \citep{Frisch95}. A similar expression stands when using the current density $j$ instead of $B$. In practice we compute the structure functions using the above PDFs of magnetic field increments, and we show them in the top row of figure~\ref{fig:DNS_vs_BxC_StructureFunctions_and_SpectrumOfExponents}. In this figure it appears that in the range of lags considered here the seven first structure functions do behave as power laws. We find that we do not need to invoque Extended Self Similarity, which consists in considering structure functions as functions of the third order structure function $S_3$, rather than of the lag in order to widen the power-law-behaving range. Hence, we have $S_n \propto \ell^{\zeta_n}$, where $\zeta_n$ is called the spectrum of exponents (in fact in the following we will normalize it with the third exponent, i.e. we will consider $\zeta_n/\zeta_3$). The dependence on $n$ of $\zeta_n$ quantifies the intermittency, as the field is intermittent if and only if $\zeta_n$ depends non-linearly on $n$. The spectra of exponents deduced from our DNS and $B \times C$ magnetic fields are shown in the bottom row of figure~\ref{fig:DNS_vs_BxC_StructureFunctions_and_SpectrumOfExponents}. As they should, they strongly deviate from Kolmogorov's linear scaling. From these plots it appears once more that both magnetic fields are non-Gaussian and that they have similar statistical properties.

\subsection{ Comparison 5: Partial Variance of Increments - Correlating intermittent current sheets with discontinuities in magnetic fields }

\begin{figure}
\includegraphics[scale=0.35]{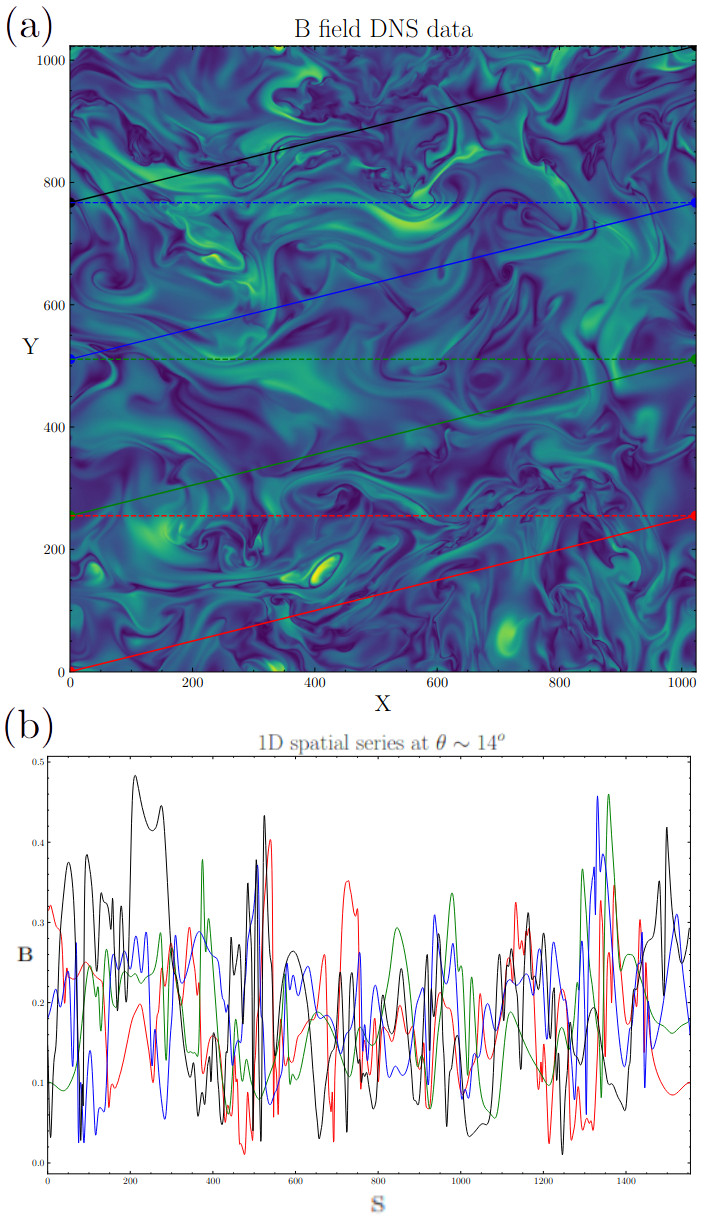}
\caption{\label{fig:1D_data_path} (Color online) (a) An example of our 1D data path in a X-Y cut used in a PVI quantification, and (b) the locally sampled data series below as function of path parameter $s$.}
\end{figure}

Our Fig.~\ref{fig:DNS_vs_BxC_PdfsOfIncrements_B_field_and_j_field} quantified increments in norms of magnetic field and current density, further used in Fig.~\ref{fig:DNS_vs_BxC_StructureFunctions_and_SpectrumOfExponents} for structure function analysis. Now, we will use similar incremental magnetic field changes along a parametrised path (using path parameter $s$) written as $\Delta \textbf{B}(s, \Delta s) = \textbf{B}(s + \Delta s) - \textbf{B}(s)$ and current density $\Delta \textbf{j}(s, \Delta s) = \textbf{j}(s + \Delta s) - \textbf{j}(s)$ to produce a statistical analysis to identify intermittent turbulent structures (i.e. current sheets) by analysing the discontinuities present in magnetic fields. This time, we measure the normalized partial variance of increments (PVI)
\begin{equation}
    I_{s,\Delta s}= \frac{|\Delta \textbf{B}(s, \Delta s)|^2}{\sqrt{\langle |\Delta \textbf{B} |\rangle}} \,,
\end{equation}
where $<\cdot> = (1/l) \int_l\cdot ds$ denotes a spatial average over the entire length $l$ of the path considered (concatenated paths across the domain), and $\Delta s$ is the spatial lag. The square of the above quantity $I_{s,\Delta s}$ is referred to as PVI as given in \cite{2008GeoRL..3519111G, 2018SSRv..214....1G}. We follow the idea presented in \cite{2011JGRA..116.9102S} to detect discontinuities along a tangential 1D path traced within 2D simulation cuts along the X, Y, and Z direction of the 3D simulation data. As shown in the top of Fig.~ \ref{fig:1D_data_path}, we sample the simulation along this 1D path which is $14^o$ with respect to X, Y and Z direction for Z, X and Y cuts, respectively. The periodicity of the data enables us to consider the entire 1D path along the domain, where the path re-enters the opposite periodic side. We have adopted this $14^o$ angle such that the offset distance between the path re-entry is greater than the integral scale of the data. Along this 1D path, we measure the partial variance of increments (PVI). This shows the correlation between current structures formed due to the turbulence and intermittent PVI events along each cut for the data sets. The PVI events for each separate tangential path on the sampled DNS data set is shown in the bottom of Fig.~\ref{fig:1D_data_path}.

\begin{figure*}
       \centering
\begin{subfigure}[b]{0.48\textwidth}
\includegraphics[width=\textwidth]{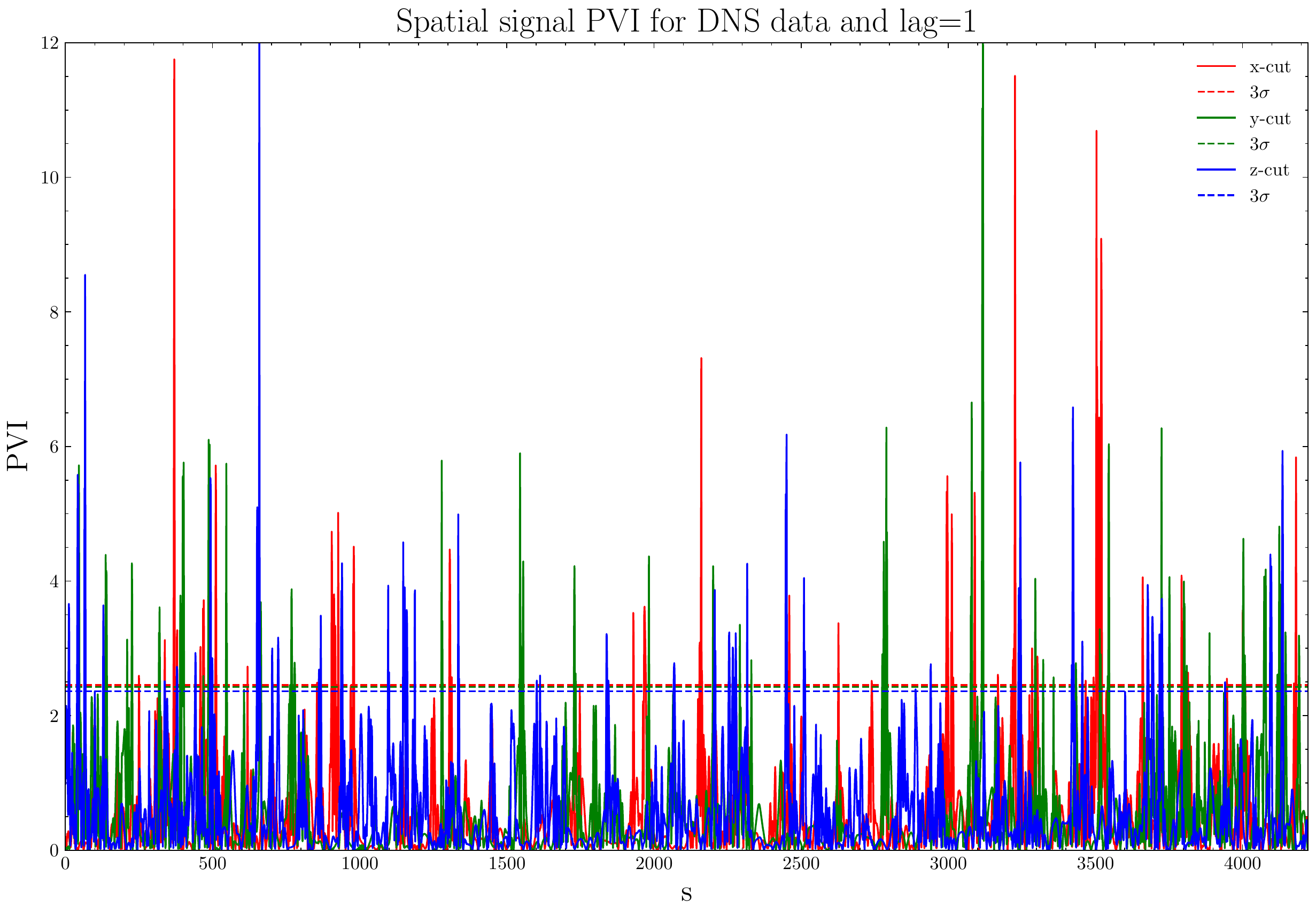}
\includegraphics[width=\textwidth]{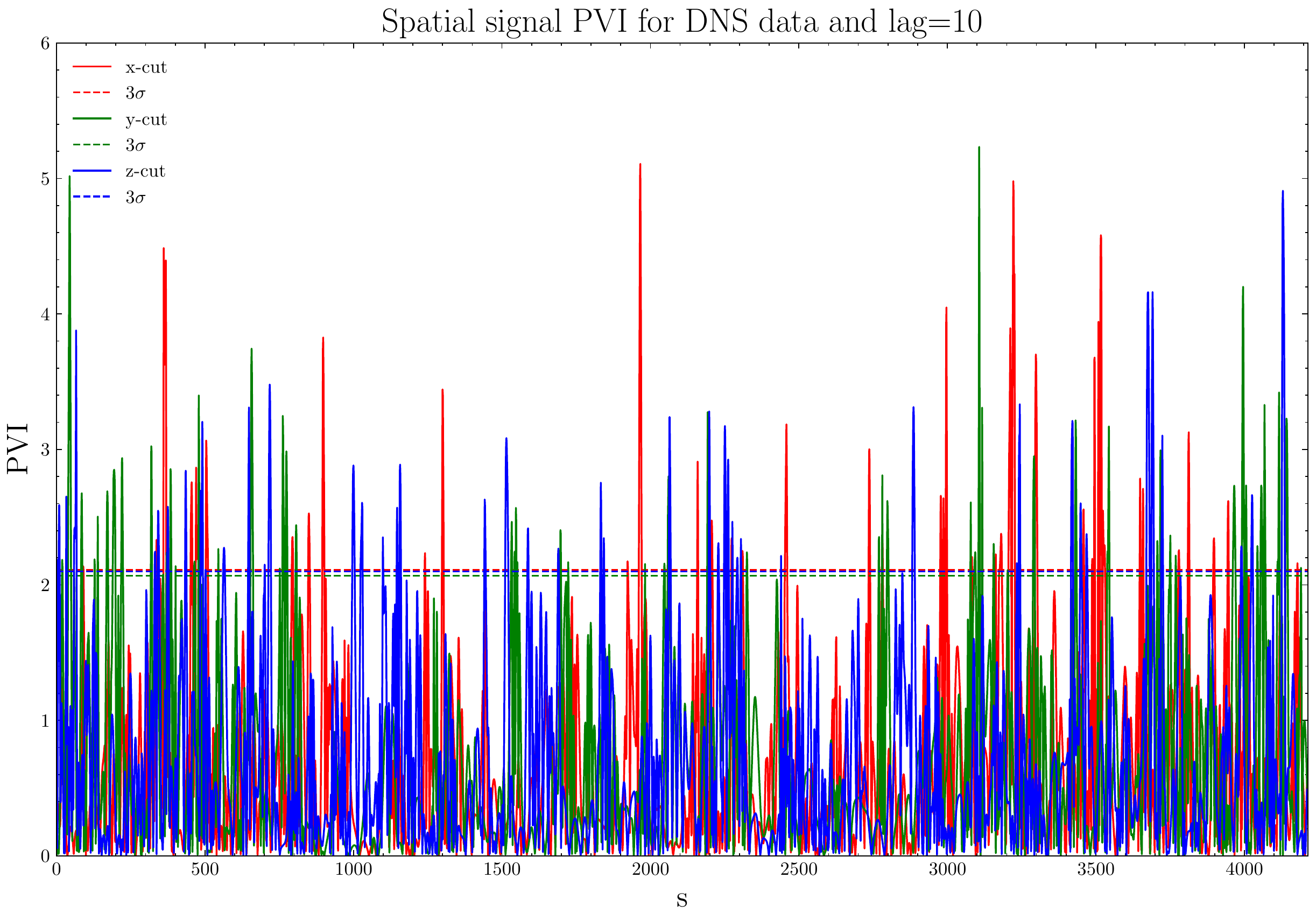}
\includegraphics[width=\textwidth]{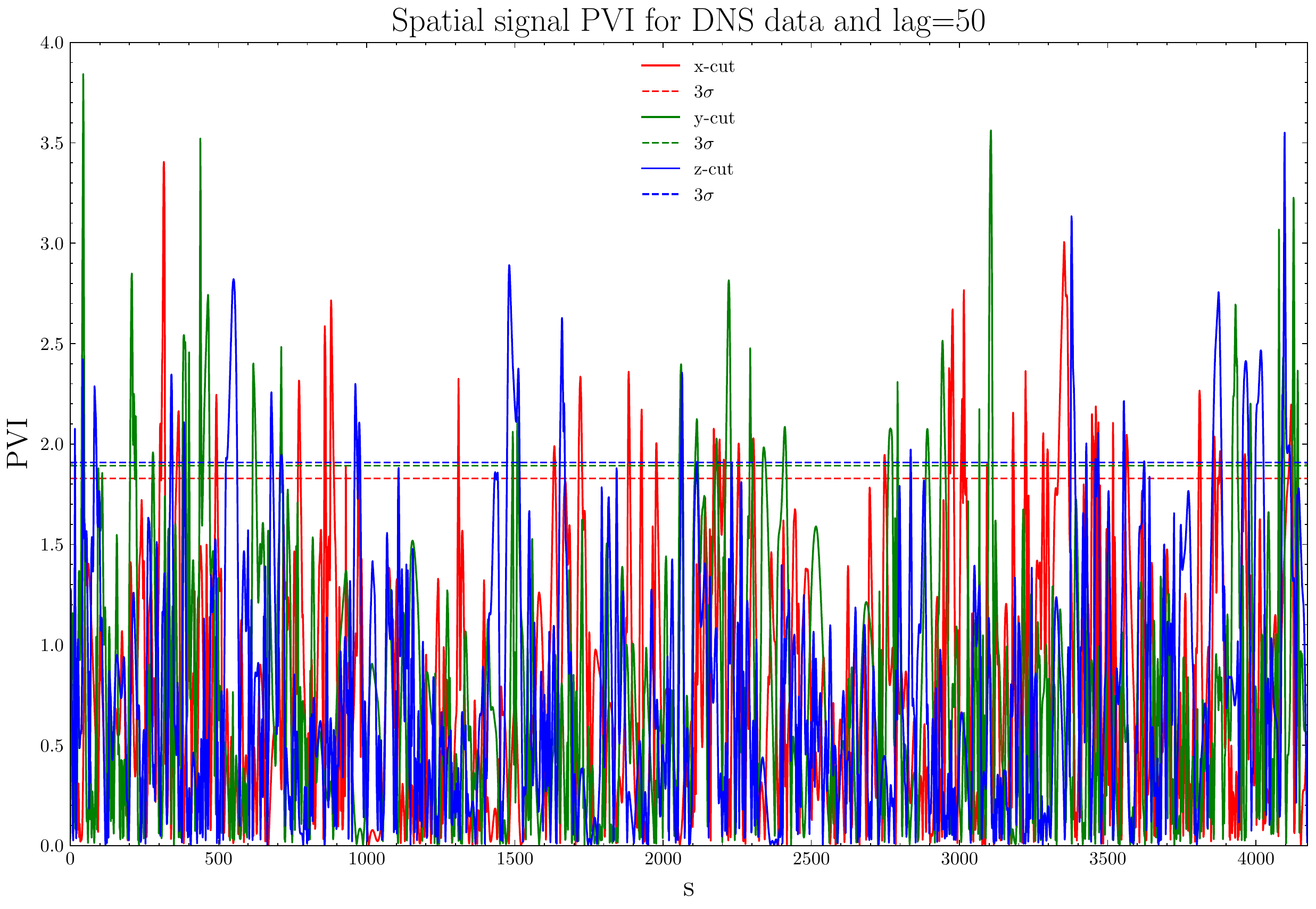}
\caption{PVI signal for DNS data}
\end{subfigure}
\begin{subfigure}[b]{0.48\textwidth}
\includegraphics[width=\textwidth]{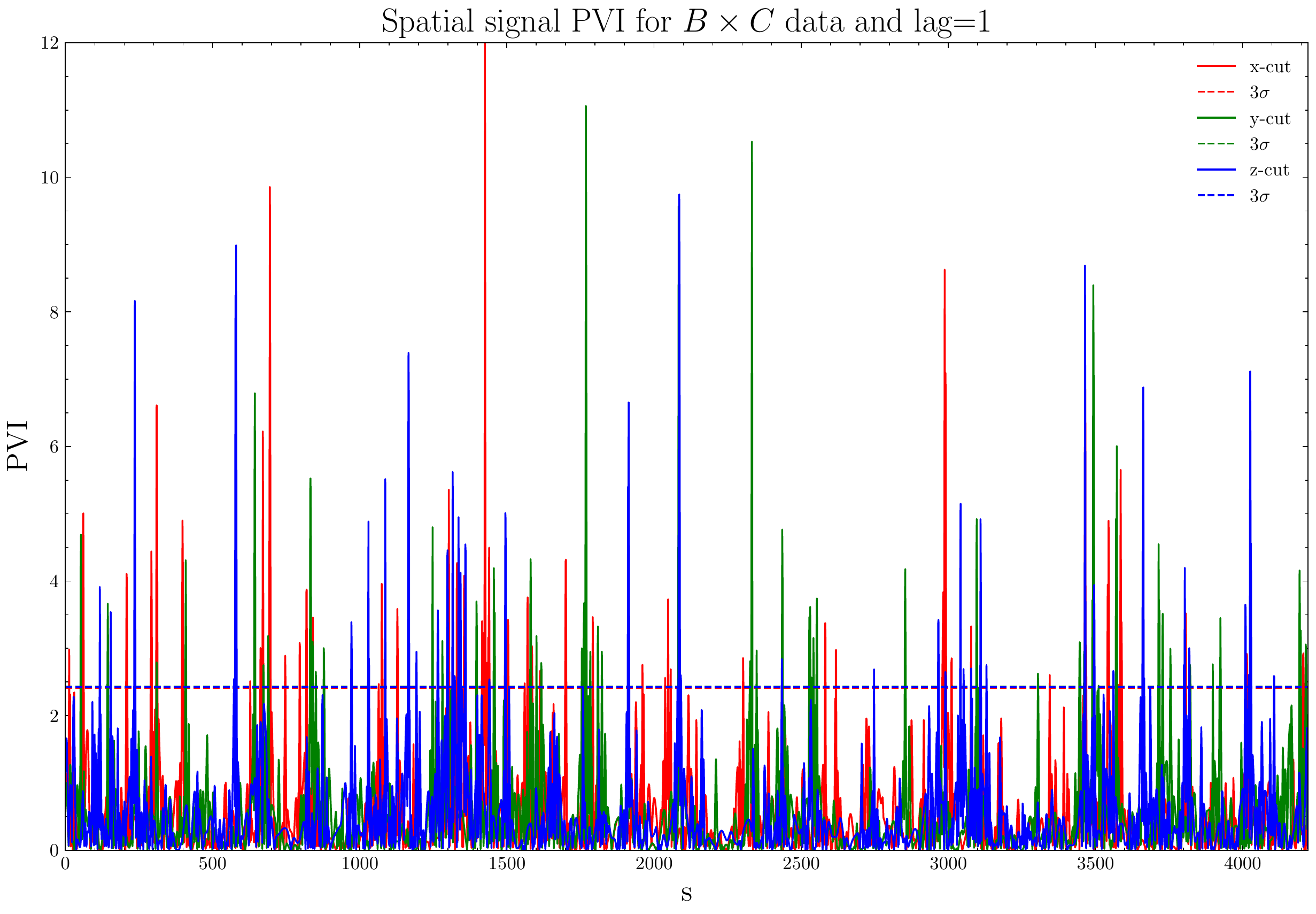}
\includegraphics[width=\textwidth]{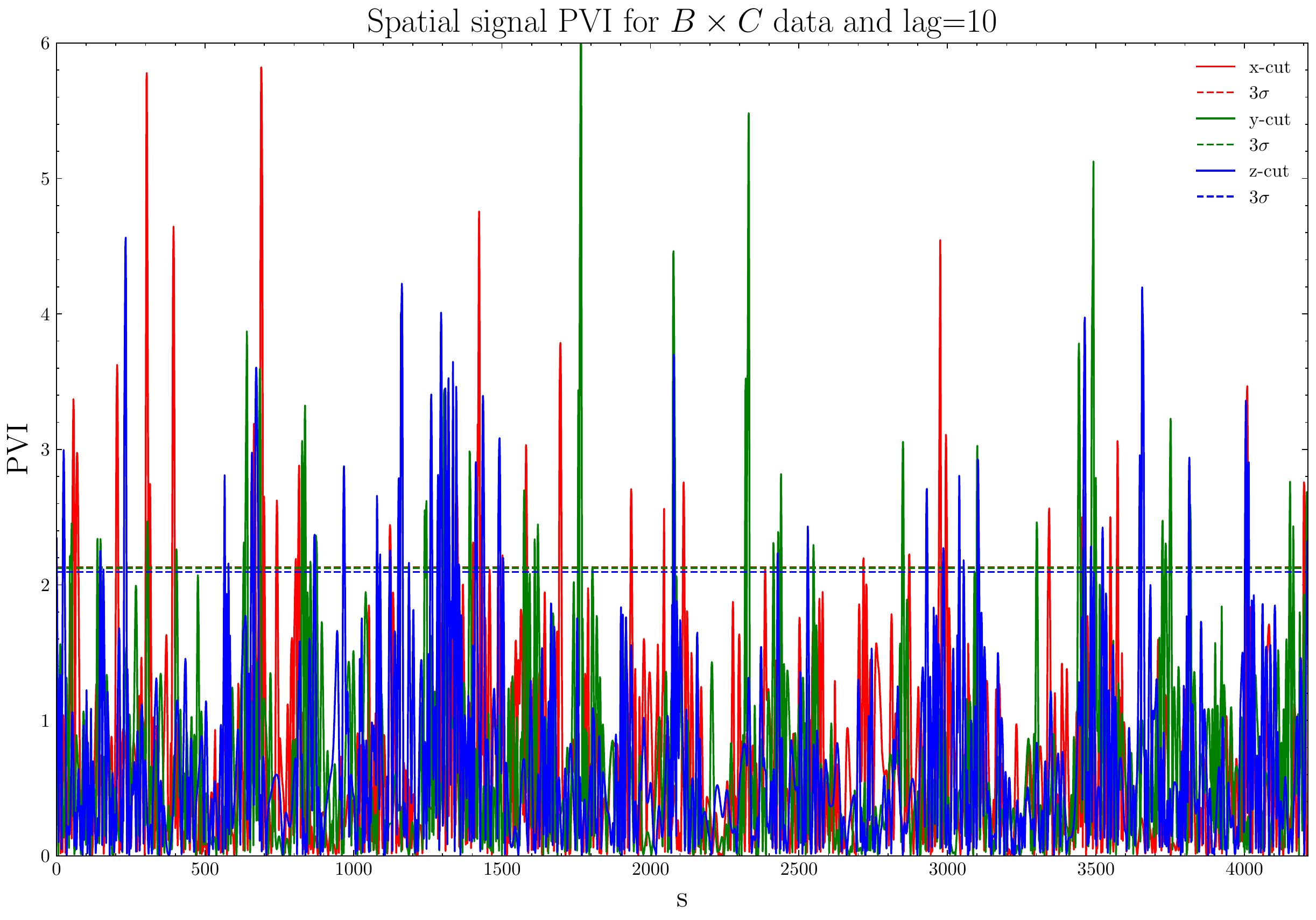}
\includegraphics[width=\textwidth]{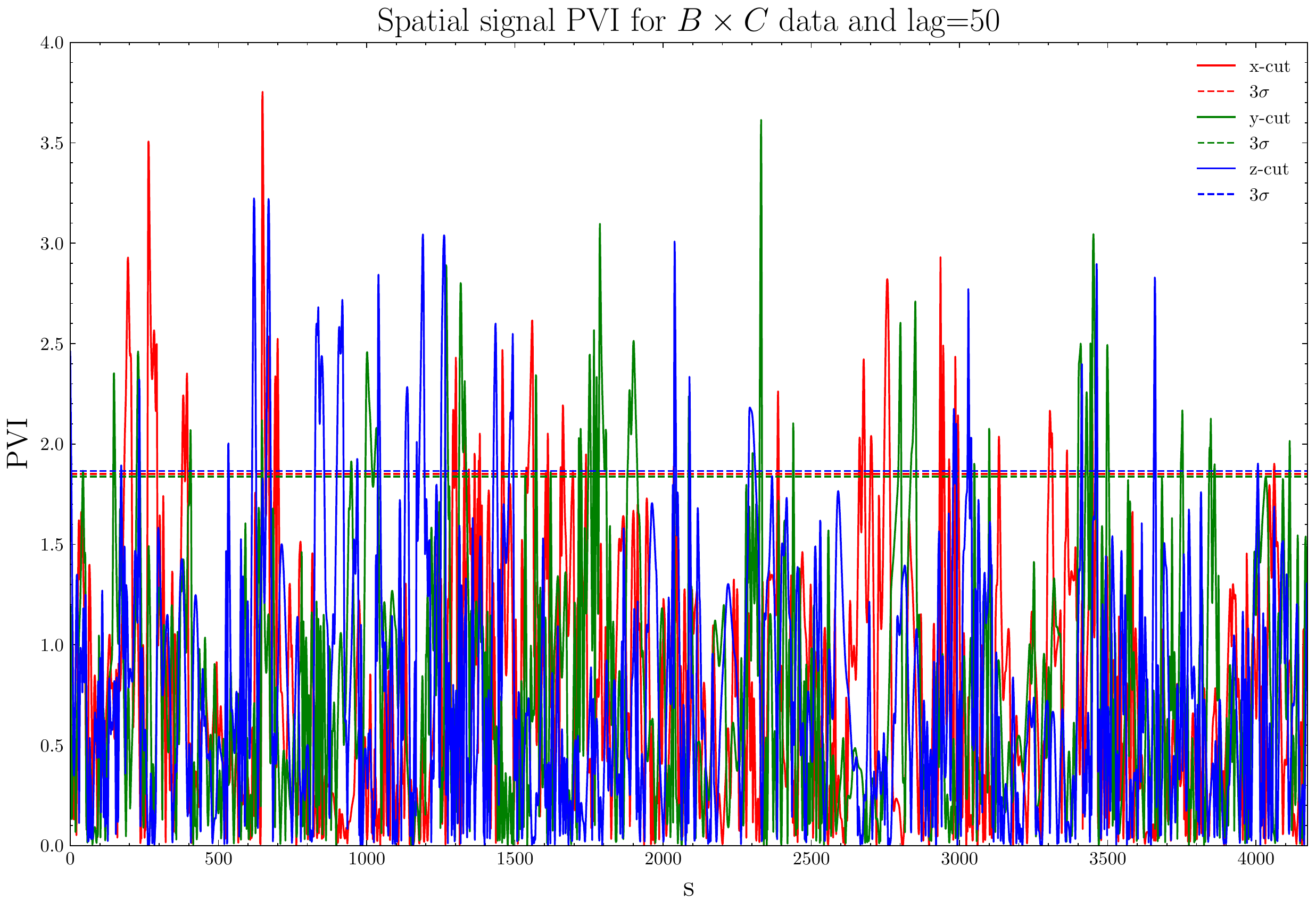}
\caption{PVI signal for $B\times C$ data}
\end{subfigure}

\caption{\label{fig:PVI w threshold} (Color online) PVIs computed for $\Delta s=1,10,50$ (top, middle, bottom). The dashed horizontal lines indicated for each cut (marked in the same color) represent the threshold of PVI $\theta = 3 \sigma$, calculated separately for each PVI series.}
    \end{figure*}

For lags $\Delta s=1,10,50$, the PVI series calculated for the DNS and $B \times C$ are plotted for X, Y and Z cuts in Fig.~\ref{fig:PVI w threshold}. The PVI series can easily measure the presence of intermittent events relating to current sheets or magnetic reconnection. In a turbulent flow, the non-Gaussian events fill up the space in addition to these very rare intermittent events, whose values lie above the standard deviation of the sample. By applying a threshold method to the PVI analysis of numerical simulations, \citep{2011JGRA..116.9102S,2013JGRA..118.4033D} found a direct correlation between PVI events satisfying the threshold parameters to the non-Gaussian and intermittent events of a flow. In our analysis, the threshold parameter $\theta$ is set to $3 \sigma$, where $\sigma$ is the standard deviation calculated across the PVI series. The increment of the threshold parameter leads to separating even higher intermittent events from the sample. We find the distinct regions of intermittent (rarely occurring) and non-Gaussian events (frequently occurring) in the turbulent flow above and below this threshold, respectively. In Fig.~\ref{fig:PVI w threshold} the PVI signal for DNS ($B \times C$) data is plotted on the left (right) column for values of different lag, $\Delta s=1,10,50$. Considering a threshold of $PVI > \theta$, the smallest lag of $\Delta s=1$ captures the highest intermittent events compared to higher lags for each cuts of the two cases. $PVI > \theta$ captures both intermittent and non-Gaussian events as we increase the lag and as such the information gets saturated with lower intermittent events which we see in all the plots. As shown in Fig.~\ref{fig:PVI w threshold}, we clearly expect the $B \times C$ to provide similar information about the discontinuities present in the magnetic field compared to the DNS data.

\begin{figure*}
\centering
\begin{subfigure}[b]{0.48\textwidth}
\includegraphics[width=\textwidth]{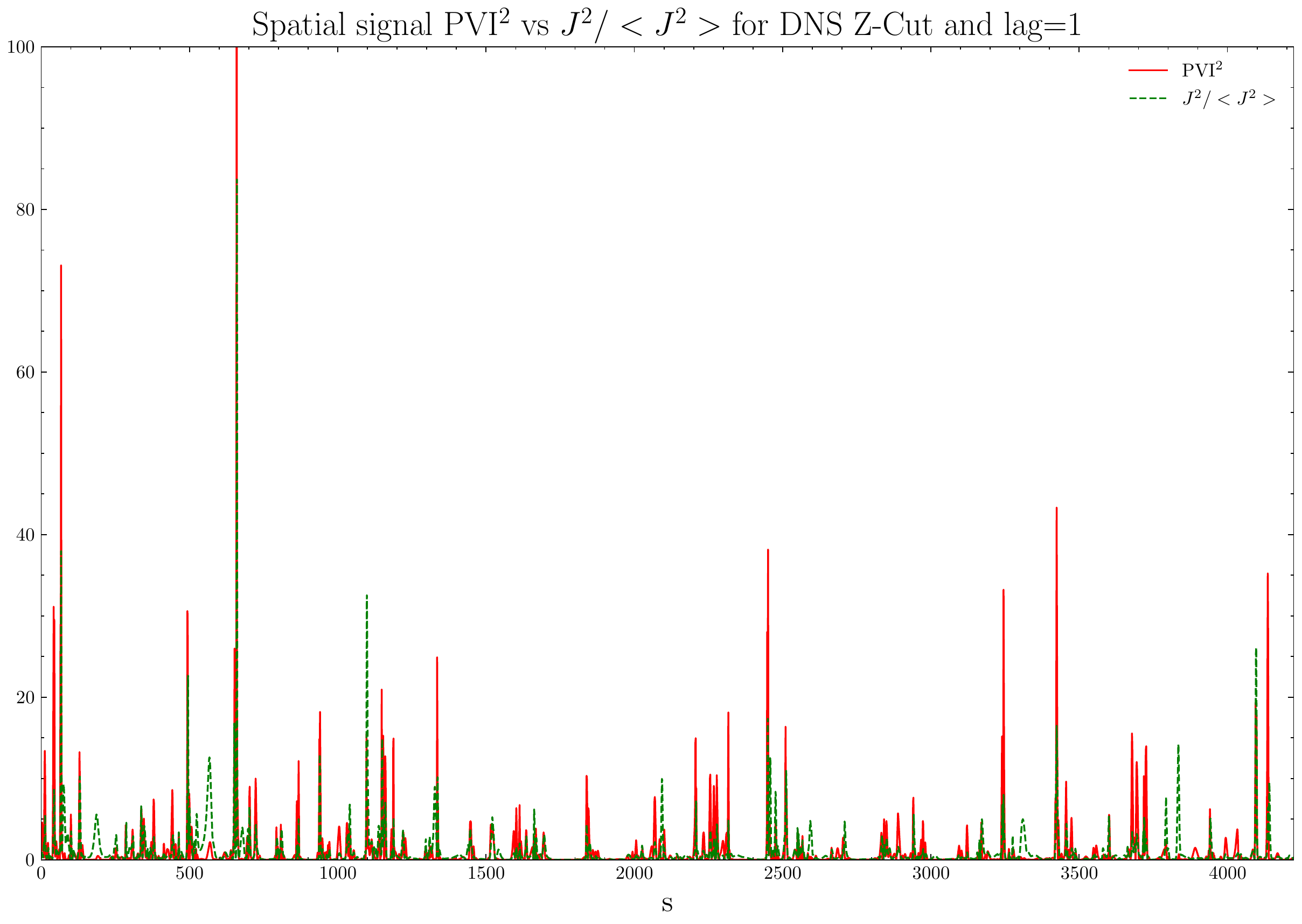}
\caption{}
\end{subfigure}
\begin{subfigure}[b]{0.48\textwidth}
\includegraphics[width=\textwidth]{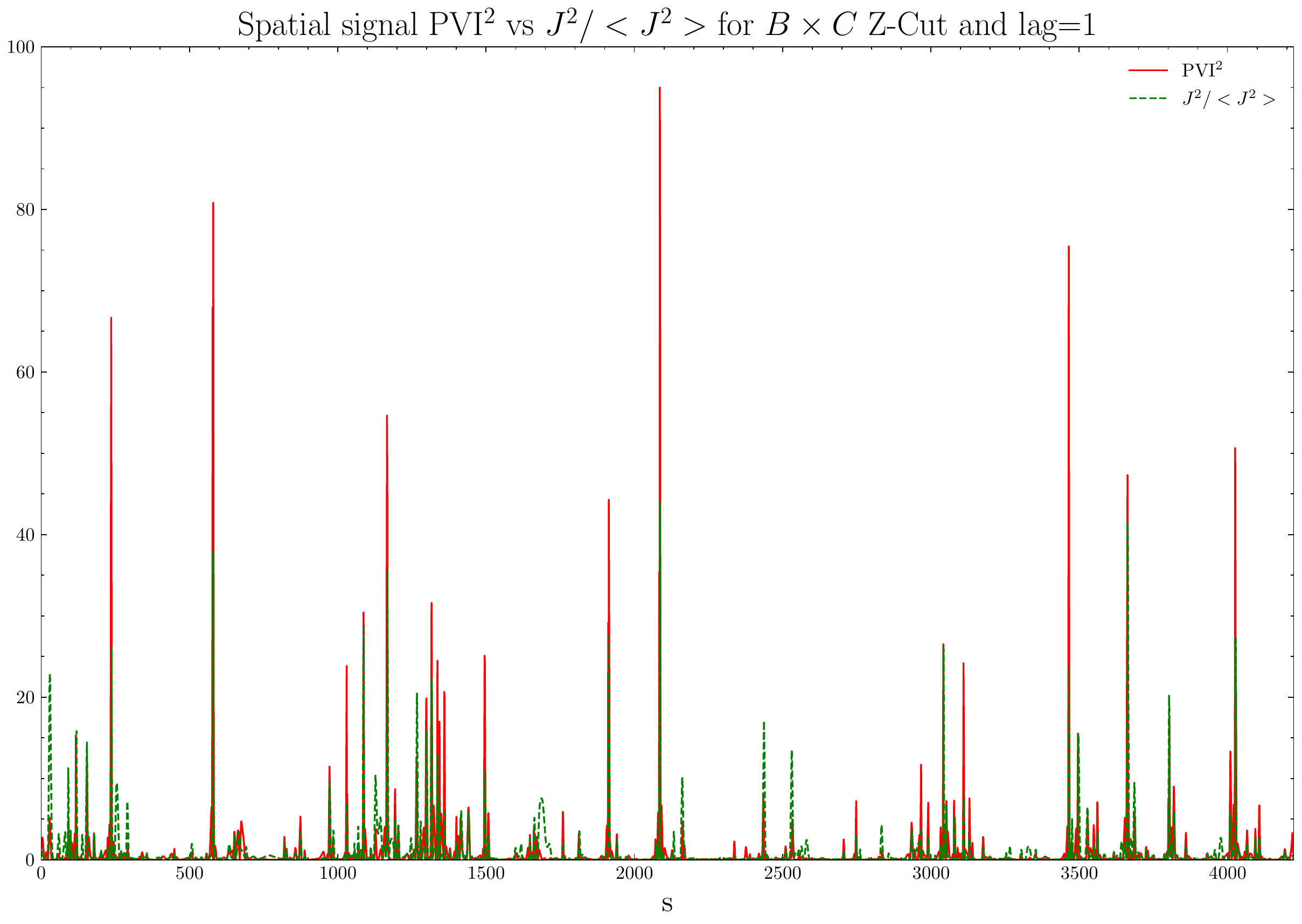}
\caption{}
\end{subfigure}
\caption{\label{fig:PVI2 vs J2} (Color online) PVI$^2$ and the square of the \textbf{J} normalized to its mean value is plotted as a spatial signal with lag $\Delta s=1$ for the Z-cut in (a) the DNS data, and (b) the $B \times C$ data.}
\end{figure*}

\begin{figure*}
\centering
\includegraphics[width=\textwidth]{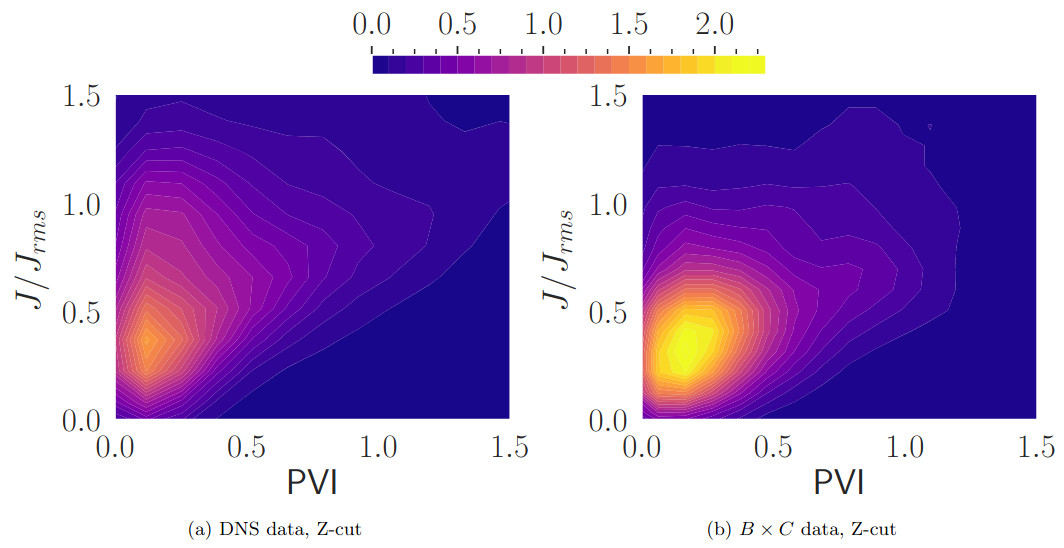}
\caption{\label{fig:joint pdf} (Color online) Kernel density estimate of the joint PDF of the magnitude of the current $\textbf{J}/ \textbf{J$_{rms}$}$  and the PVI values for the detected PVI events. The PVI signal has been computed on a spatial separation $\Delta s =1$ and the Pearson correlation coefficient is 0.62 for DNS (a) and 0.71 for $B \times C$ (b)}
\end{figure*}

As a follow-up study, we present our analysis to correlate the intermittent events found by analyzing the increments of the magnetic field to the presence of magnetic reconnection or current sheet events. According to \cite{2018SSRv..214....1G}, the more prominent peaks of current density correspond statistically to more significant peaks of PVI. It is because of this that the PVI method can describe and identify the strong magnetic gradients. We compare the spatial signals of PVI$^2$ (in red) to $\textbf{J}^2/<\textbf{J}^2>$ (in dashed green) for Z-cut in Fig.~\ref{fig:PVI2 vs J2}. We analyze both the signals for a lag of $\Delta s=1$. The reference DNS data shows distinct statistical peaks to be in phase of the PVI signal and the current density. As so, it demonstrates that the two quantities have a positive cross-correlation. The same can be interpreted for the $B \times C$ data, which presents similar statistical results in Fig.~\ref{fig:PVI2 vs J2}, demonstrating that the PVI method in this case is capable of successfully relating the magnetic field discontinuities to estimate the intermittency in current density for the $B \times C$. A further statistical study to show the relation of PVI and current is shown by the joint PDF in Fig.~\ref{fig:joint pdf}. We plot the kernel density estimate of the joint PDF for PVI value compared to the $\textbf{J}/ \textbf{J$_{rms}$}$ at the smallest spatial separation of $\Delta s=1$ for the Z-cut of DNS (Fig.~\ref{fig:joint pdf}(a)) and $B \times C$ (Fig.~\ref{fig:joint pdf}(b)). For both cases, a positive correlation is seen with the extreme values of PVI corresponding to the extreme values of current density and the bulk of the PVI population at lower PVI values corresponds to the lower current density values. The Pearson correlation coefficient between the variables is 0.62 for the DNS data and 0.71 for the $B \times C$ data. Thus, it shows how PVI helps in identifying these extreme events from magnetic discontinuities and in doing so relates them to the sharp gradients in current density effectively for $B \times C$ as it should be for the DNS data.

\section{Outlook}

We introduced a novel, geometrically controlled, parametrized way to generate mock turbulent MHD fields, emphasizing the magnetic field and current density variations in 3D space as typically encountered in high resolution DNS data. Our $B\times C$ model was shown to correspond visually, as well as statistically, with typical isotropic turbulent magnetic fields. In contrast to DNS models, our tool is not computationally intensive, and has direct parametric control on the spectral properties embedded in these turbulent fields. By generalizing this proof-of-concept to cases with also background organized fields, our model may become a direct tool for testing rivaling MHD (anisotropic) turbulence theories, and for inspecting their visual appearance. Potential applications of this tool are numerous \footnote{For instance, to explore the correlation between magnetic fields and cosmic rays propagation in the interstellar medium, one has to run costly numerical simulations \citep{SetaEtAl18}, while with the present model one can generate the necessary prescribed fields much faster (and thus generate more of them, to improve the statistics), controlling their statistical properties and thus quantify the sensitivity of the propagation to these properties.}, with the distinct advantage that laptop resources suffice. This can then quickly generate turbulent magnetic data cubes, to study e.g., polarized light propagation through astrophysical turbulent media (Faraday effect); or for fitting our geometric parameters to match actual 3D DNS fields, that can then be artificially `upscaled' to ever larger sizes $N^3$. Note that we can likewise generate pure hydro fields, where vorticity-velocity vectors behave like our current-magnetic vector fields, and hence produce data cubes for both incompressible flow and magnetic field vectors, for input to full MHD simulations with particular turbulent properties. Future work can try to generate a suitable generalization of this model for isotropic MHD turbulent fields, to those encountered in situations with a clear organized guide field, where differences in behaviour parallel versus perpendicular to the guide field can be explored.

\section{Acknowledgements}

We thank the referees for their constructive comments. RK and JBD are supported by Internal funds KU Leuven, project C14/19/089 TRACESpace. RK further received funding from the European Research Council (ERC) under the European Union's Horizon 2020 research and innovation programme (grant agreement no. 833251 PROMINENT ERC-ADG 2018) and a FWO project G0B4521N. PL acknowledges support from the European Research Council, under the European Community's Seventh framework Programme, through the Advanced Grant MIST (FP7/2017-2022, No 742719).

\bibliography{BxC_2022}

\end{document}